\documentclass[a4paper,11pt]{article}
\pdfoutput=1
\usepackage{jheppub}
\usepackage[utf8]{inputenc}
\usepackage{amsmath}
\usepackage{amssymb}
\title{\boldmath A Study of confinement  for $Q\bar{Q}$
potentials on D3, M2 \& M5 branes}

\author{Edward Quijada and Henrique Boschi-Filho}

\affiliation{Instituto de F\'\i sica, Universidade Federal de Rio de Janeiro,\\Caixa Postal 68528, RJ 21941-972, Brazil}

\emailAdd{edward@if.ufrj.br}
\emailAdd{boschi@if.ufrj.br}

\abstract{We study analytically and numerically the interaction potentials between a pair of 
quark an anti-quark on D3, M2 and M5 branes.
These potentials are obtained using  Maldacena's method involving Wilson loops 
and present confining and non-confining behaviours in different situations that we explore in this work.
In particular, at the near horizon geometry the potentials are non-confining in agreement with conformal field theory expectations. 
On the other side, far from horizon, the dual field theories are no longer conformal and the potentials present confinement. 
This is in agreement with the behaviour of strings in flat space where the string mimics the expected flux tube of QCD. 
A study of the transition between the confining/non-confining regimes in the three different scenarios (D3, M2, M5) is also performed.}

\begin{document}
 \maketitle
\flushbottom
\section{Introduction}

Usually, in quantum field theory, the Wilson loop operator is defined as $$W(C)=\dfrac{1}{N}TrPe^{i\oint_{C}A},$$ 
where $C$ denotes a closed loop in space-time and the trace is over
the fundamental representation of the gauge field $A$ with $SU(N)$ symmetry. 
In the particular case of a rectangular 
loop (of sides $T$ and $L$), it is possible to calculate 
(in the limit $T\rightarrow\infty$) the expectation value for the Wilson loop: 
$$<W(C)>={\cal A}(L)e^{-TE(L)},$$ where $E(L)$ can be identified with the energy of the quark-antiquark pair in the static limit.

Soon after the conjecture about the duality between M/string theory in $AdS$ spaces and conformal gauge 
field theories \cite{maldacena1,GKP,Witten1,Review1,Review2}, 
Maldacena \cite{maldacena2}, Rey and Yee \cite{Rey:1998ik} (MRY), 
proposed a method to calculate expectation values of the Wilson loop for the large $N$ limit of field theories.  
This limit is calculated from a string theory in a given background using the gauge/gravity duality. 
 
In this method, the expectation value of the Wilson loop is related 
to the worldsheet area $S$ of a string whose boundary is the loop in 
question such that $$<W(C)>\sim e^{S}.$$

Maldacena  used this approach to calculate the quark-antiquark potential for the  string in the 
$AdS_5\times S^5$ background \cite{maldacena2} obtaining a non-confining potential for the 
infinitely massive quark-antiquark pair, consistent with the conformal symmetry of the dual super Yang-Mills theory. 
In other backgrounds the quark-antiquark potential can be confining as shown for instance in  
\cite{kinar}, where a confinement criterion was obtained. 

This approach can also be extended to the finite temperature case
 \cite{finiteT,brandhuber} by considering an AdS Schwarzschild background. In this case, 
 the temperature of the conformal dual theory is identified with the Hawking 
temperature of the black hole \cite{Witten2}. 
This situation also leads to a non-confining potential for the quark-antiquark 
interaction.

The thermodynamic of D-brane probes in a black hole background were treated in \cite{Mateos:2007vn}. These systems are  holographically dual
to a small number of flavours in a finite-temperature gauge theory. First order phase transitions were found characterised by a  
confinement/deconfinement  transition of quarks.

A phenomenological approach was also considered calculating the Wilson loop for the string in some 
holographic AdS/QCD models. For instance, the hard-wall model exhibits a confining behaviour 
\cite{henrique2,Andreev:2006ct} reproducing the Cornell potential.
At finite temperature, this calculation gives a second order phase transition describing qualitatively a 
confinement/deconfinement phase transition \cite{henrique3}. 
Then, it was shown that a Hawking-Page phase transition \cite{HP} should occur for the hard- and 
soft-wall models at finite temperature 
\cite{Herzog,henrique4,Andreev:2006eh,Andreev:2006nw,Kajantie:2006hv}. In particular, 
for the soft-wall model, an interesting estimate of the deconfinement temperature was found \cite{Herzog}, 
compatible with QCD expectations. 

In a recent  paper  it were studied some geometric configurations of a static string on a D3-brane background \cite{henrique} and  
also a string-like object on  M2- and M5-brane backgrounds \cite{Quijada:2015tma}. These  geometric configurations corresponds to a gauge 
theory which describes the quark-antiquark interaction on the branes. For some specific geodesic regimes we found confining interactions and for 
others non-confining potentials were found. 

In this paper we perform a systematic analytical and numerical study of the quark-antiquark potentials in D3- M2- and M5- brane backgrounds analysing their confining/non-confining behaviours in different situations, always at zero temperature. 
In particular, at the near horizon geometry the potentials are non-confining in agreement with conformal field theory expectations. 
On the other side, far from horizon, the dual field theories are no longer conformal and the potentials present confinement. 
This is in agreement with the expected behaviour of strings in flat space where the string mimics the flux tube model of QCD. In the cases of M2 and M5 branes in M-theory we choose a cigar-shaped membrane background such that that stringy picture of the dual flux tube also holds. We also focus in searching for the point in the geodesics at which the zero temperature confinement/deconfinement transition takes place. 
  
  \section{Wilson loops in D3- M2- and M5-brane spaces}
 
\begin{figure}[h]
 \centering
 \includegraphics[width=10cm,height=6cm]{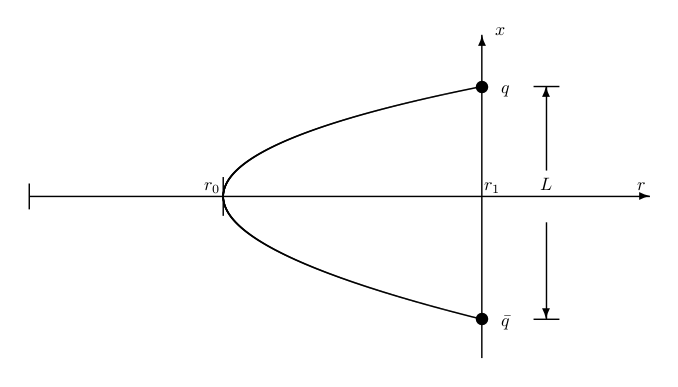}
 \caption{Position of quark $q$ and anti-quark $\bar q$ on the D3-brane (represented here by the $x$-axis) 
 together with the static string as the curve which connects $q$ and $\bar q$ through $r_0$.}
 \label{fig:one}
\end{figure}

\begin{figure}[h]
 \centering
 \includegraphics[width=10cm,height=6cm]{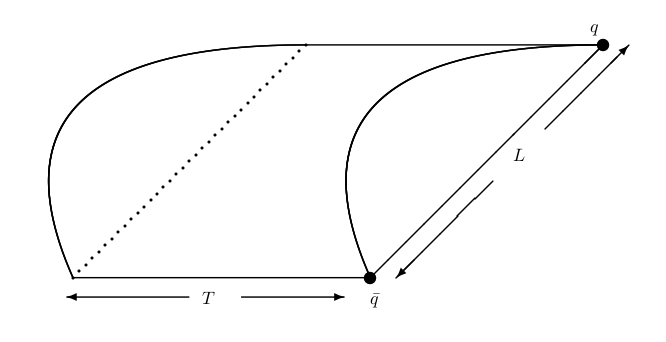}
 \caption{Curved worldsheet surface of the static string corresponding to a time interval $T$. The associated Wilson loop is the plane rectangle of sides $T$ and $L$.}
 \label{fig:two}
\end{figure}

We start this study by considering the Wilson loop on the background generated by a large number of coincident D3-branes in string theory in 10 dimensional spacetime. 
The Nambu-Goto string action  \cite{Becker:2007zj}: 
\begin{equation}
 S=\frac{1}{2\pi}\int d\sigma d\tau\sqrt{\det(g_{NM}\partial_{\alpha}X^N\partial_{\beta}X^M}) 
\end{equation}
is employed  on this background, where the scale was set to $\alpha'=1$, $X^N(\sigma,\tau)$ are the coordinates of the string worldsheet and $g_{NM}$ is the background metric. The specific form of the metric is given in the next section. 
It is considered that the  pair of quark-antiquark is contained in the D3-brane world which are attached to the ends of the open string 
that lives in  10 dimensions. For simplicity, we work in a static string configuration, that is represented in 
figure \ref{fig:one}. The Wilson loop corresponds to a rectangle  with sides $L$ and $T$, where $T$ is some time interval. This
rectangle is associated with the worldsheet surface as shown in figure \ref{fig:two}.

Thus the distance separation $L$ between the quark-antiquark pair may be computed starting from the geodesic of the static string on this 
D3-brane background. This distance turns out to be an expression in terms of $r_0$ and $r_1$ (respectively, minimum and maximum for $r$ coordinate in the worldsheet. See \cite{kinar, henrique}):
\begin{equation} \label{L}
 L=2\int \frac{g(r)}{f(r)}\frac{f(r_0)}{\sqrt{f^2(r)-f^2(r_0)}}dr,
\end{equation}
where 
\begin{align*}
f^2(r)&=(2\pi)^{-2}g_{00}(r)g_{ii}(r)\,,\\
g^2(r)&=(2\pi)^{-2}g_{00}(r)g_{rr}(r)\,.
\end{align*}

According to the MRY proposal the worldsheet area ($S$) is proportional to the energy interaction ($V$) between the 
 quark-antiquark pair, so it may  also  be written down in terms of $r_0$ and $r_1$: 
\begin{equation}\label{V}
V=2\int_{r_0}^{r_1}\frac{g(r)f(r)}{\sqrt{f^2(r)-f^2(r_0)}}dr-2m_q \,.
\end{equation}
Note that it is in general necessary to subtract the masses of the quarks $m_q$ in order to obtain a finite result for the energy interaction.

We continue our study  analysing the cases concerning  M2- and M5-brane backgrounds. Since these backgrounds of 11-dimensional SUGRA corresponds to M-theory objects, it is not possible  to start from Nambu-Goto action. Instead we should start from a 11-dimensional membrane action in those backgrounds \cite{Duff:1990xz, Becker:2007zj}:
\begin{eqnarray}
S=\frac{1}{(2\pi)^2l_{11}^3}\int d^3\sigma\Big(\frac{(-\gamma)^{1/2}}{2}\left[\gamma^{ij}\partial_i X^{M}\partial_j X^{N}
G_{MN}(X)-1\right] \nonumber\\
 +\epsilon^{ijk}\partial_i X^{M}\partial_j X^{N}\partial_k X^{P}A_{MNP}(X)\Big)\,,
\end{eqnarray}
where $i,j=0,1,2$ are world-volume indices with $\gamma^{ij}$ as the induced metric, $M,N,P=0,...10$  are space-time
indices with $G_{NM}$ as the space-time metric, $X^N(\sigma^0,\sigma^1,\sigma^2)$ are the membrane coordinates, $A_{MNP}$ is 
a three-form field with with strength $F=dA$ and $l_{11}$ sets the scale for the membrane (see \cite{Duff:1990xz}).
After compactification of one spatial dimension of the membrane wrapped along the 11-th dimension of space-time we are able to reduce 
the membrane in 11 dimensions to a string-like object in 10 dimensions (see \cite{duff}). As a result we are able to work with string-like objects 
and similarly to the case of strings on D3 backgrounds, we utilize the static configuration and the MRY proposal to get 
the distance separation and  energy interaction between a pair of quark-antiquark on M2-(M5-)branes (see \cite{Quijada:2015tma}).
  

\section{D3-brane}

The solitonic solution of 10-dimensional supergravity that we are going to study is a space geometry generated by $N$ coincident D3-branes. This solution is usually written down as \cite{Horowitz:1991cd,GKP}: 
\begin{equation}\label{0}
 ds^2 =\left(1+\frac{R^4}{r^4}\right)^{-1/2}(-dt^2 + dx_3^2) +\left(1+\frac{R^4}{r^4}\right)^{1/2}(dr^2 + r^2 d\Omega^2_5)\,,
\end{equation}
where $R$ is a constant defined by $ R^4 = 4\pi g Nl_s^4 $.

Following the MRY approach, the calculation of the distance separation $L$, Eq. \eqref{L}, and the static potential interaction $V$, Eq. \eqref{V},  
between a pair of quarks on the  D3-brane were obtained in \cite{henrique}:
 \begin{equation}\label{1}
  L=\frac{2r_0^3}{R^2}I_1\left(\frac{r_1}{r_0}\right)+\frac{2R^2}{r_0}I_2\left(\frac{r_1}{r_0}\right)\,,
 \end{equation}
 \begin{equation}\label{2}
  V=\frac{2r_0\sqrt{r_0^4+R^4}}{2\pi R^2l_s^2}I_1\left(\frac{r_1}{r_0}\right)-2m_q\,,
 \end{equation}
 where 
 \begin{equation}\label{2.1}
  I_1\left(\frac{r_1}{r_0}\right)=\int_1^{r_1/r_0}dy\frac{y^2}{\sqrt{y^4-1}}\,,
 \end{equation}
\begin{equation}\label{2.2}
 I_2\left(\frac{r_1}{r_0}\right)=\int_1^{r_1/r_0}dy\frac{1}{y^2\sqrt{y^4-1}}\,.
\end{equation}
\noindent 
Following \cite{kinar} we have that the  quark mass must be:
\begin{equation}
 2m_q= \frac{r_1}{\pi l_s^2}\,, 
\end{equation}
which diverges in the limit $r_1 \to \infty$. 

In the following we are going to study the distance separation $L$, Eq. \eqref{1},  and the potential energy $V$, Eq. \eqref{2},  of the quark-antiquark pair in various different situations in the D3-brane solution. 

\subsection{Non-confining behaviour}

\begin{figure}[h]
 \centering
 \includegraphics[width=7cm,height=5cm]{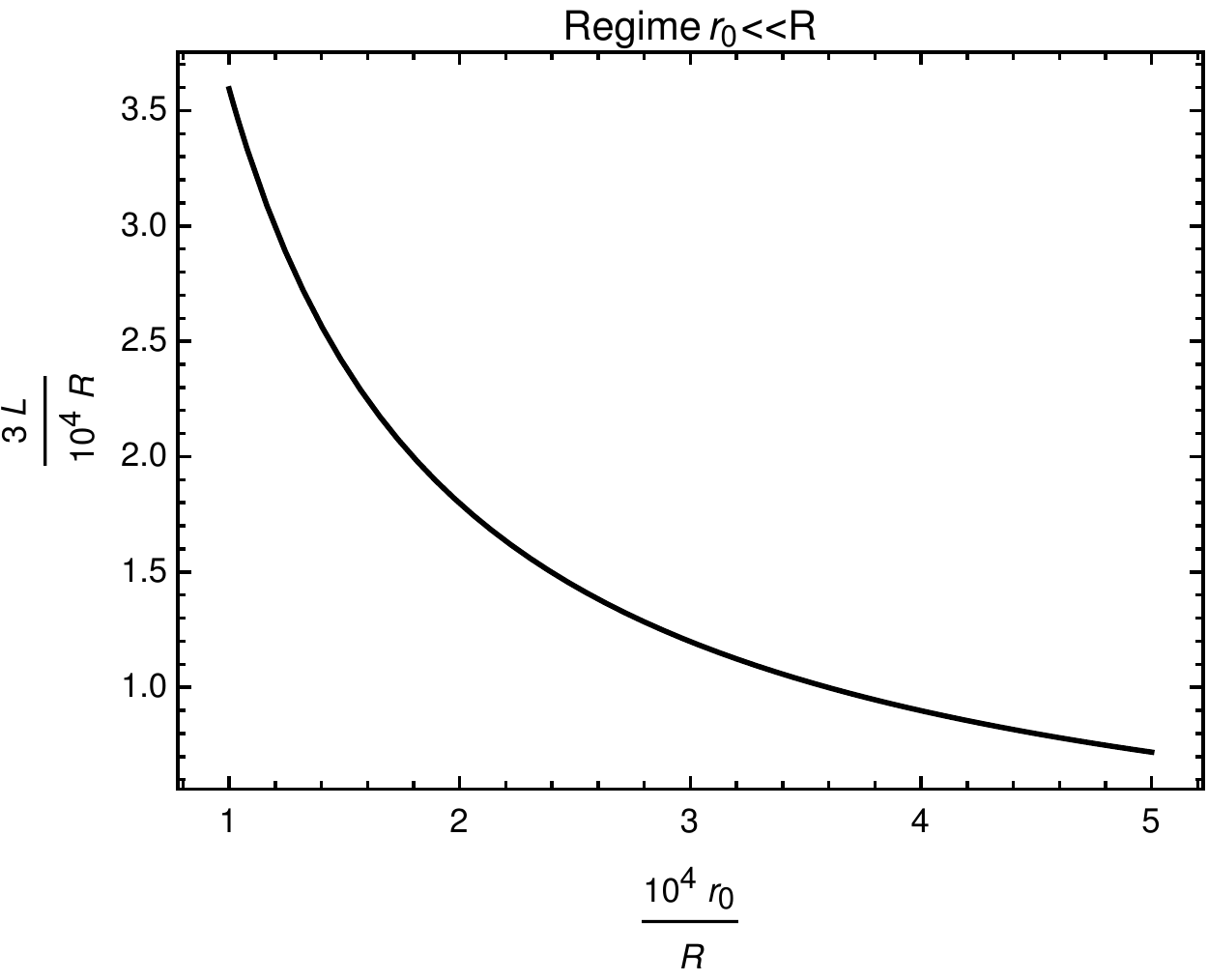}
 \caption{Monotonic decreasing behaviour of $L/R$ vs. $r_0/R$. Here $r_1/R=10^5$ and $r_0/R<5 \times 10^{-4}$. }
 \label{fig3}
\end{figure}

Let us start our study considering the regime defined by $r_1>>r_0$ which means that the quark is very massive. Also we take $r_0<<R$
which means that we are in the near horizon geometry which corresponds approximately to the AdS$_5$ space.  
We take $r_0$ as the independent variable of parametrization with fixed $R$. Then, the behaviour of the distance separation $L$, Eq. \eqref{1}, against $r_0$ is analysed. The numerical result is shown in figure \ref{fig3}, where we plot $L/R$ vs. $r_0/R$. This plot shows a monotonic decreasing behaviour of $L/R$ against  $r_0/R$. 

\begin{figure}[h]
 \centering
 \includegraphics[width=7cm,height=5cm]{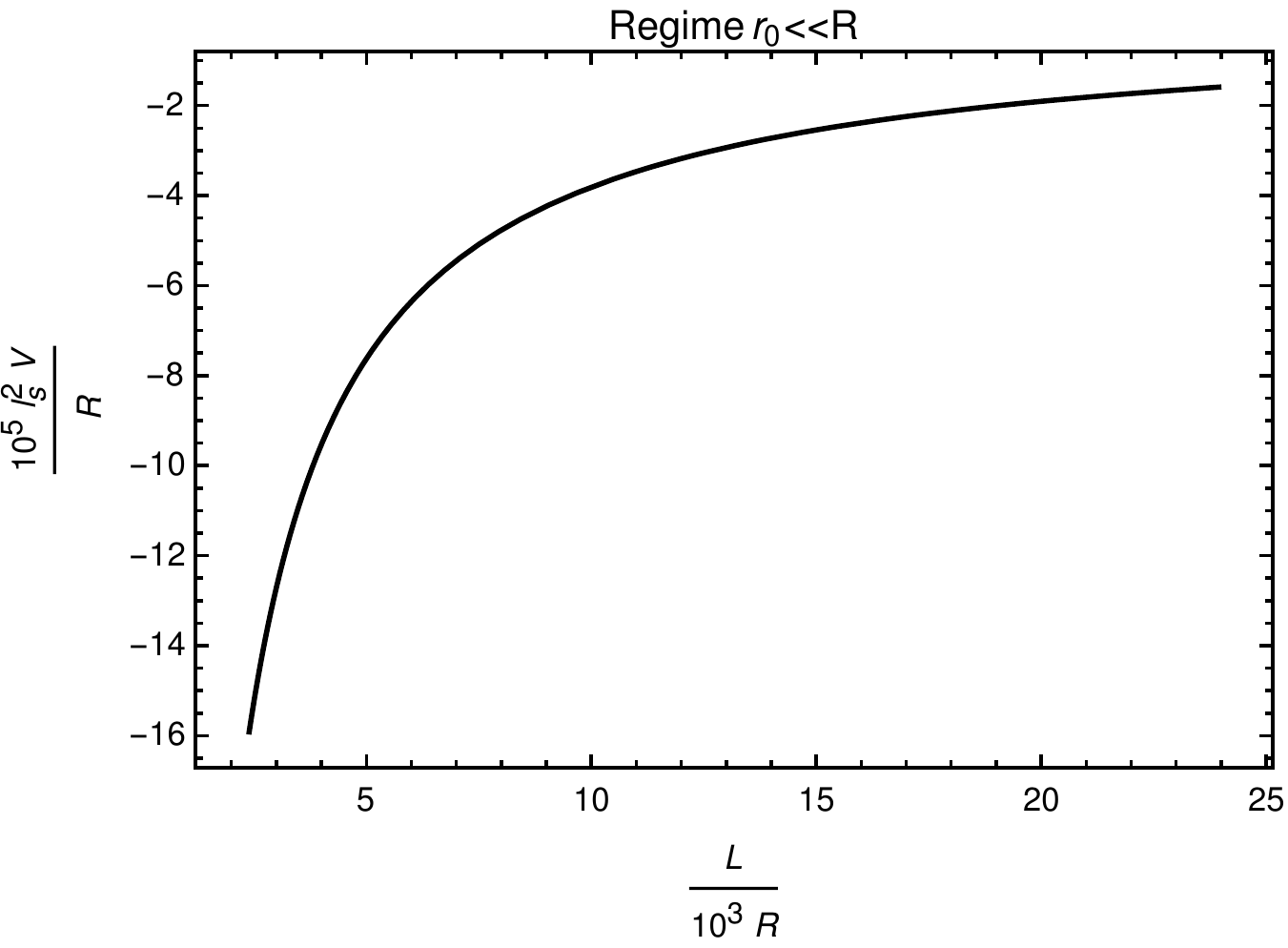}
 \caption{Non-confining potential interaction $l_s^2V/R $ vs. $L/R$. Here $r_1/R=10^5$ and $r_0/R<5 \times 10^{-4}$.}
 \label{fig4}
\end{figure}

The next step is to analyse the behaviour of the potential $V$, Eq. \eqref{2}, against the separation $L$, Eq. \eqref{1}. The numerical result is shown in figure \ref{fig4}, where 
we plot  $V/R$ versus $L/R$. 
This plot shows an increasing function which goes to zero as $L$ increases. So one can conclude that this plot corresponds to a non-confining  potential which is essentially Coulomb like, as the one found by Maldacena in \cite{maldacena2} for the case of the pure AdS space. This result is also in agreement with 
\cite{henrique} where a non-confining potential $V\sim-1/L$  was obtained in the regime $r_1>>r_0$ with $r_0<<R$. The dual field theory in this case is the well known $\cal N$=4 SYM which is a superconformal field theory. Then the non-confining behaviour found for the Wilson loop is in agreement with the conformal property of the dual theory. 

\subsection{Confining behaviour}

\begin{figure}[h]
 \centering
 \includegraphics[width=7cm,height=6cm]{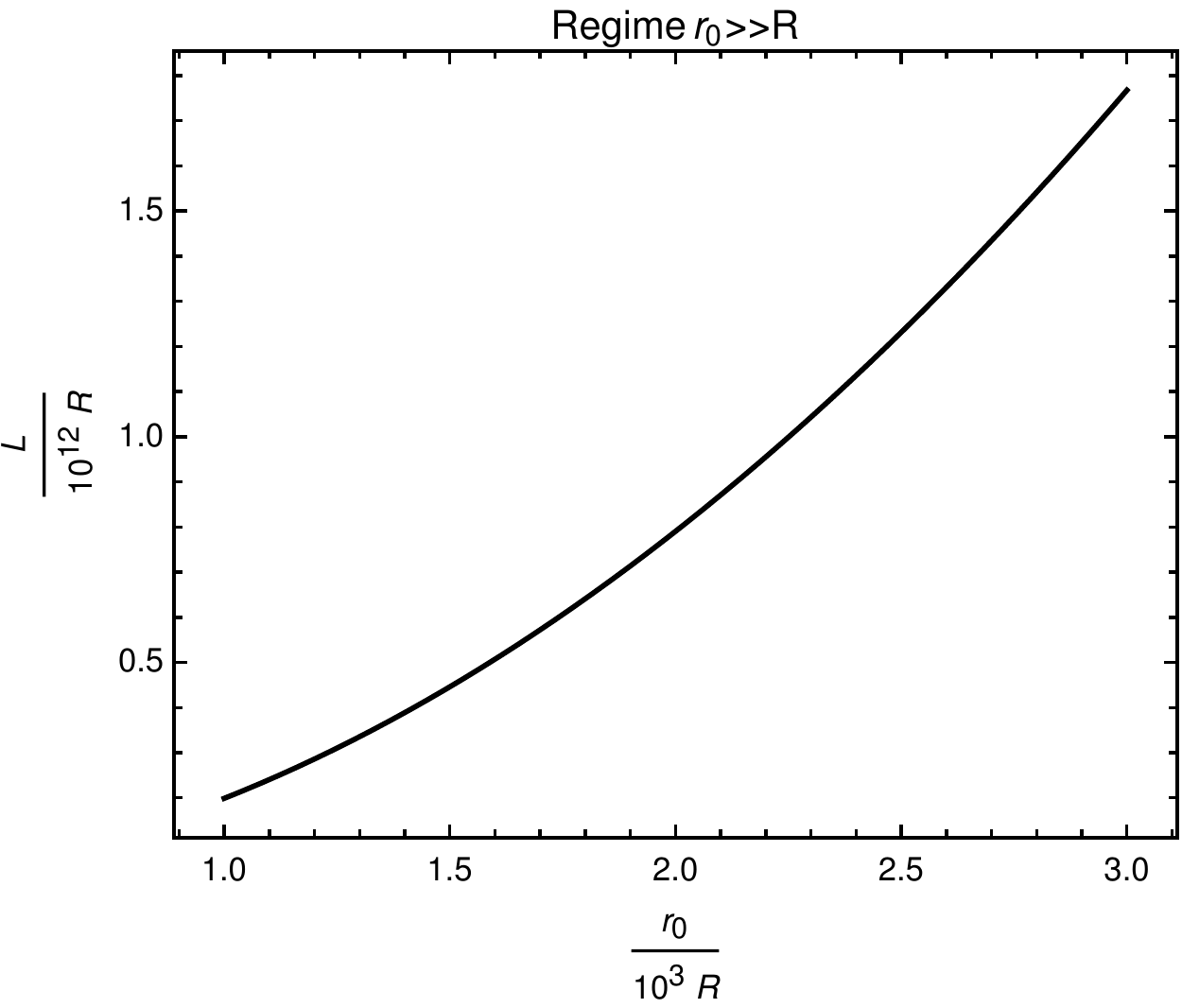}
 \caption{Increasing behaviour of $L/R$ vs. $r_0/R$. Here $r_1/R=10^5$ and $r_0/R>10^3$.}
 \label{fig5}
\end{figure}

Our next step is to analyse the regime $r_1>>r_0$ (very massive quark)  but with $r_0>>R$ which corresponds to the region far from the horizon which is approximately a flat space geometry. First we perform a numerical study of the distance separation $L$, Eq. \eqref{1}, against the minimum position of the string $r_0$. The result of this analysis is presented  in figure \ref{fig5}, where we plot  $L/R$ vs. $r_0/R$. 
This figure shows a monotonic increasing behaviour of $L/R$ against $r_0/R$.

\begin{figure}[h]
 \centering
 \includegraphics[width=6cm,height=8cm]{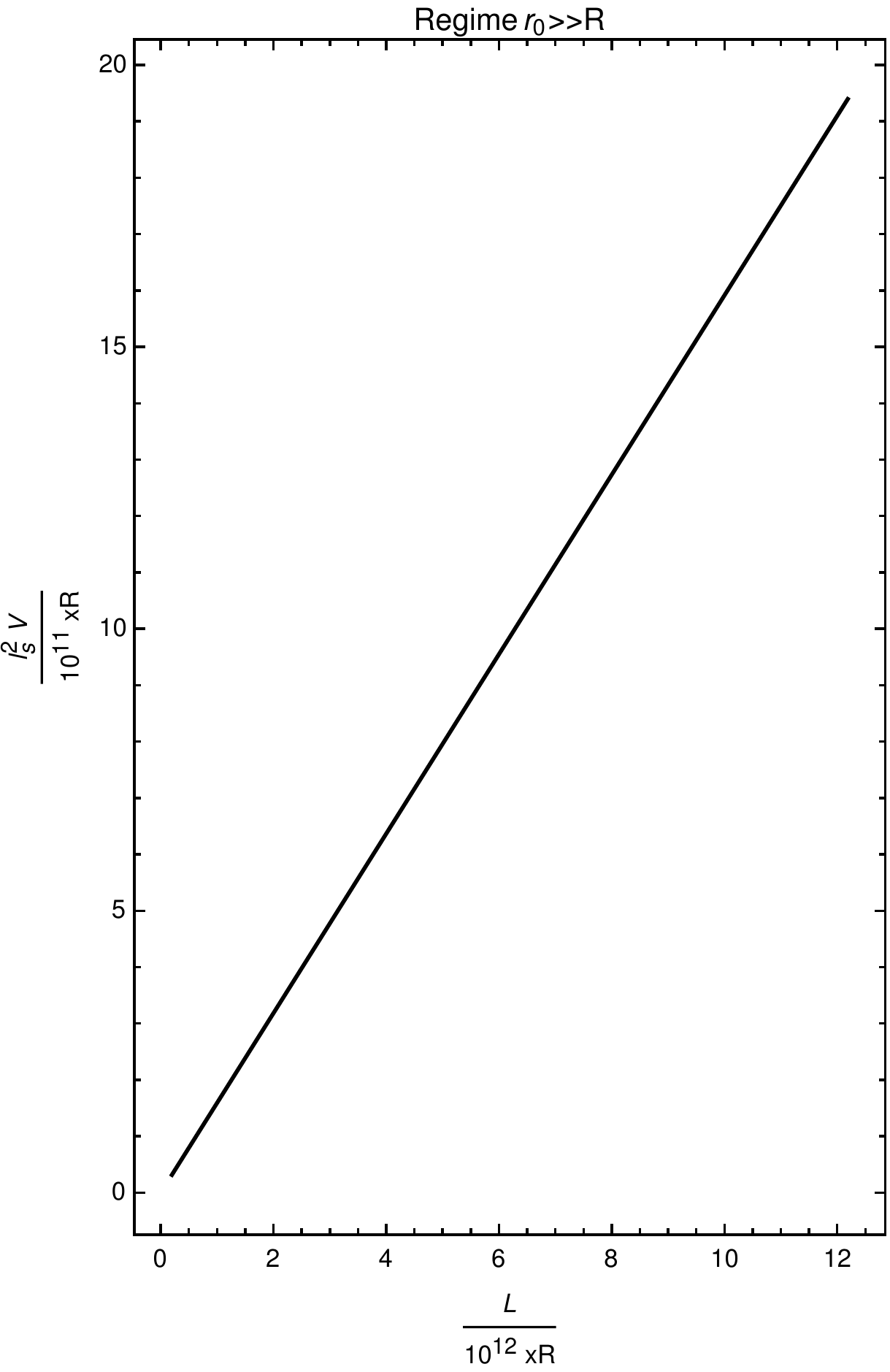}
 \caption{Plot of  $V/R$ vs. $L/R$ with $r_1/R=10^5$ and $r_0/R>10^3$ which shows a confining potential.}
 \label{fig6}
\end{figure}

Then, the next step is to study the shape of the potential $V$, Eq. \eqref{2}, against the separation distance $L$, Eq. \eqref{1}. We did this numerical study and the result is presented in figure \ref{fig6}, where 
we plot the behaviour of $V/R$ against $L/R$. 

 Looking at figure \ref{fig6} we see an almost straight line with positive derivative indicating that this 
  plot implies a confining potential. This result is in agreement with ref. \cite{henrique}, where a linear confining potential was obtained in this regime for the quark antiquark pair in D3-brane space. The dual theory in this case is no longer conformal, since we are far from the horizon. Here, we can understand this picture as a string in flat space which mimics the confining flux tube of QCD. 
 
 
\subsection{Deconfinement/Confinement transition}

In previous sections we obtained confining and non-confining behaviours for the potential energy $V$, Eq. \eqref{2}, against the separation distance $L$, Eq. \eqref{1}, in D3-brane space for different regimes of $r_0$ compared with $R$. So, we expect that a transition should occur between the regimes $r_0>>R$ (far from the horizon) and $r_0<<R$ (near the horizon).

In this section we work with $r_1>>r_0$ for  $r_0$ values in the regime $r_0\sim R$ such that we may find some deconfinement/confinement transition. Note that this is not a thermal phase transition since we are working at zero temperature. Instead, the expected transition should be related to the geometry of the D3-brane space. 

\begin{figure}[h]
 \centering
 \includegraphics[width=8cm,height=5cm]{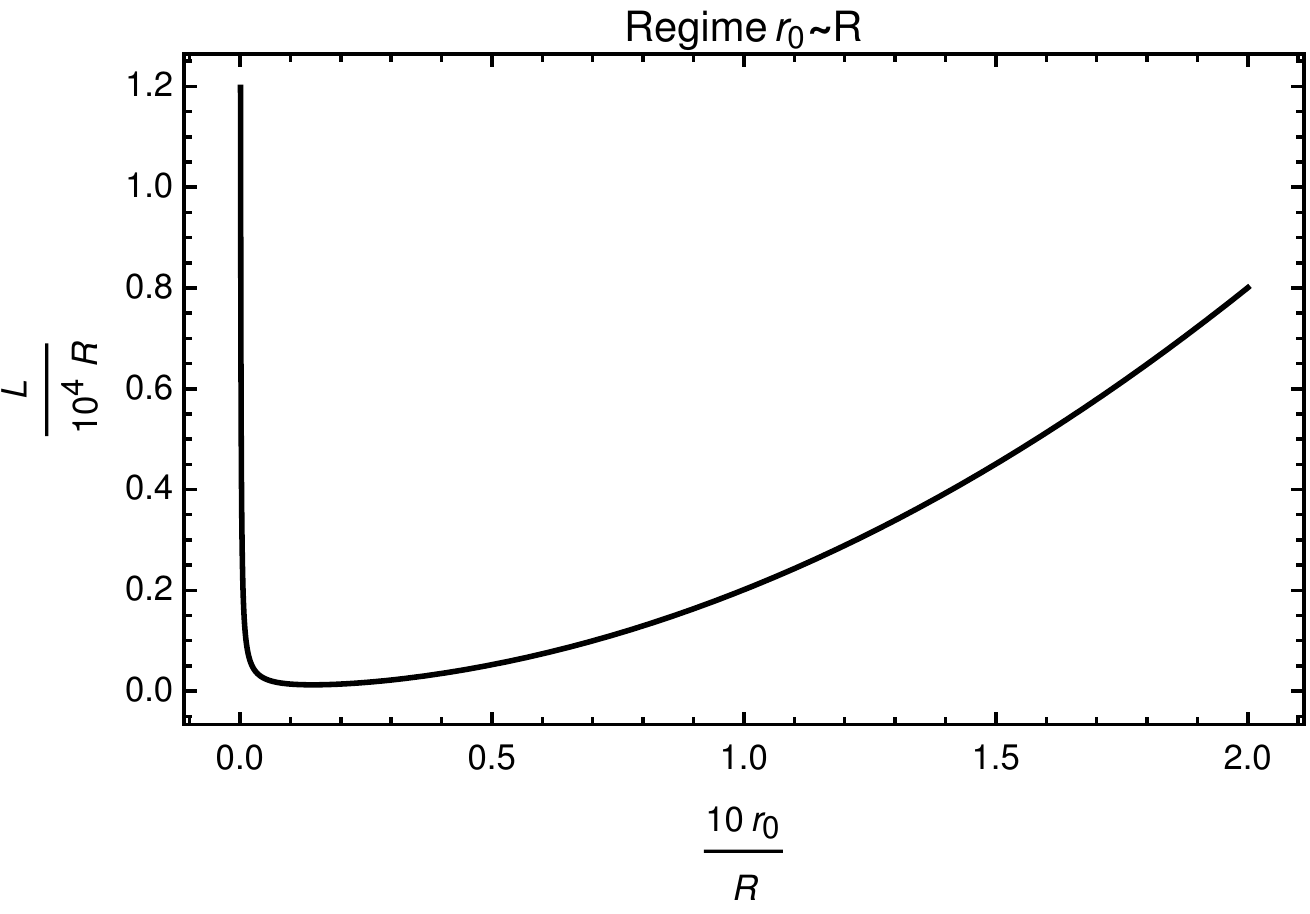}
 \caption{ $L/R$ vs. $r_0/R$. Here $r_1/R=10^5$ and $r_0/R<0.2$ .} 
 \label{fig7}
\end{figure}

First we present in figure \ref{fig7} a plot showing how $L/R$ varies against $r_0/R$. This picture shows a minimum value of $r_0$ which we call $r*$. 

Note that it is also possible to define $r*$  
from equations (\ref{1}), (\ref{2.1}) and (\ref{2.2}). Using these equations we get an equation whose root is precisely  $r*$: 
\begin{equation}\label{r*D3}
\frac{6r*^2}{R^2}I_{1}\left(\frac{r_1}{r*}\right)-\frac{2r_1^3}{R^2r*\sqrt{(r_1/r*)^4-1}} 
-\frac{2R^2}{r*^2}I_{2}\left(\frac{r_1}{r*}\right)-\frac{2R^2}{r_1r*\sqrt{(r_1/r*)^4-1}} =0
\end{equation}
 Some numerical solutions for this equation are shown in table \ref{my-label}.
\begin{table}[h]
\centering
\caption{Some values of $r*/R$ in eq. \eqref{r*D3} for different values of $r_1/R$.}
\label{my-label}
\begin{tabular}{|l|l|ll}
\cline{1-2}
$r*/R$ & $r_1/R$ &   &  \\ \cline{1-2}
$9.87\times 10^{-3}$       & $10^6$             &  &   \\ \cline{1-2}
$ 3.015 \times 10^{-2}$       & $10^5 $            &  &  \\ \cline{1-2}
$6.687 \times 10^{-2}$       & $10^4 $           &  &  \\ \cline{1-2}
$1.42 \times 10^{-1}$       & $10^3 $         &  &  \\ \cline{1-2}
\end{tabular}
\end{table}

\begin{figure}[h]
 \centering
 \includegraphics[width=7cm,height=6cm]{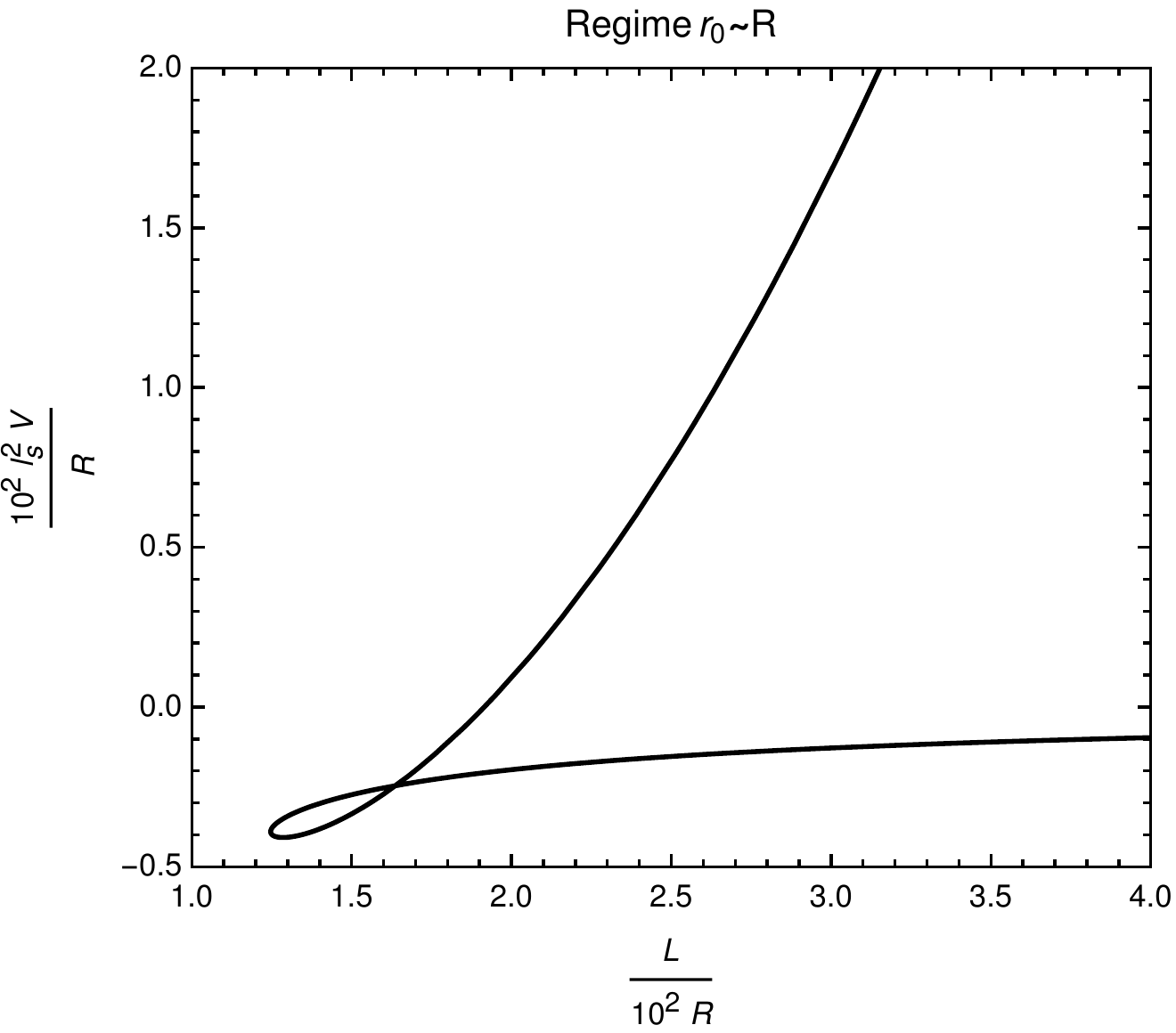}
 \caption{Transition of the potential interaction. Here $r_1/R=10^5$.} 
 \label{fig8}
\end{figure}

The potential interaction $V$, Eq. \eqref{2}, against the separation distance $L$, Eq. \eqref{1}, in this regime  is presented in  figure \ref{fig8}. In this figure we can notice
that there are two branches: the inferior one is a non-confining Coulomb-like potential, and the superior one  is a confining potential that has
a monotonic increasing behaviour as $L/R$ is increased.

From fig. \ref{fig8}, our analysis shows that  the $r_0<r*$  condition corresponds to  non-confining behaviour and $r_0>r*$ condition corresponds to confining one. This is the expected transition in the confinement/deconfinement behaviour of the quark-antiquark pair potential in D3-brane space. The transition seems to occur near the region $r_0 \sim r*$. It is important to remark that this is not a thermal phase transition since we are working at zero temperature and the transition is of geometrical nature.

\section{M2-brane}

In the previous section, we presented an analysis of the Wilson loop for the D3-brane background. Here in this section and in the following we are going to present a similar discussion for other backgrounds such as M2- and M5-brane  spaces. 
Although  these background spaces belong to 11-dimensional M-theory that must correspond to higher dimensional objects like membranes, it is possible to do a dimensional reduction. This consists in compactifying one dimension of the membrane along one spacial direction, in order to have a string-like configuration in 10-dimensional
background spaces.  For details see \cite{duff, Quijada:2015tma}.

We start the study of confinement with the MRY method in  SUGRA-backgrounds with the case of the space generated 
by $N$ coincident M2-branes.
The 11-dimensional supergravity M2-brane solution is given by the metric (see \cite{Duff:1990xz,Review2,Review1}): 
\begin{equation}
ds^2_{\text{M2}}=\left(1+\frac{R_2^6}{r^6}\right)^{-2/3}dx_3^2+\left(1+\frac{R_2^6}{r^6}\right)^{1/3}(dr^2+r^2d^2\Omega_7^2), 
\end{equation}
where $R_2$ is a constant defined by $R_2=(32\pi Nl^6_{11})^{1/6}$,
$N$ is the number of coincident branes, $l_{11}$ is the Plank's length in eleven dimensions and $d\Omega_7$ is the differential solid angle for seven-sphere.

In a previous work \cite{Quijada:2015tma}  the distance separation ($L$) and static potential (V) for a pair of quarks in a M2-brane 
space were obtained: 

 \begin{equation}\label{3.1}
L=2r_0\int_{1}^{r_1/r_0}dy\frac{\left(1+\frac{\epsilon}{y^6}\right)y^3}{\sqrt{y^8-y^6+\epsilon(y^8-1)}},
\end{equation}

\begin{equation}\label{3.2}
 V=\frac{2r_0^2\sqrt{1+\epsilon}}{2\pi l_{11}^3}\int_1^{y_1}dy\frac{y^5}{\sqrt{y^8-y^6+\epsilon(y^8-1)}}-2m_q,
\end{equation}
 where  $r_0$ (- $r_1$) is the minimum(-maximum) value of coordinate $r$  associated with the string-like object obtained by dimensional reduction and $\epsilon=(R_2/r_0)^6$. Again following ref. \cite{kinar}, 
 we can compute   the quark mass  $m_q$ as:
\begin{equation}
2m_q=\frac{r_1^2}{2\pi l_{11}^3}\,,
\end{equation}
which diverges if we let $r_1 \to \infty$. 

\subsection{Non-confining behaviour}

\begin{figure}[h]
 \centering
 \includegraphics[width=7cm,height=6cm]{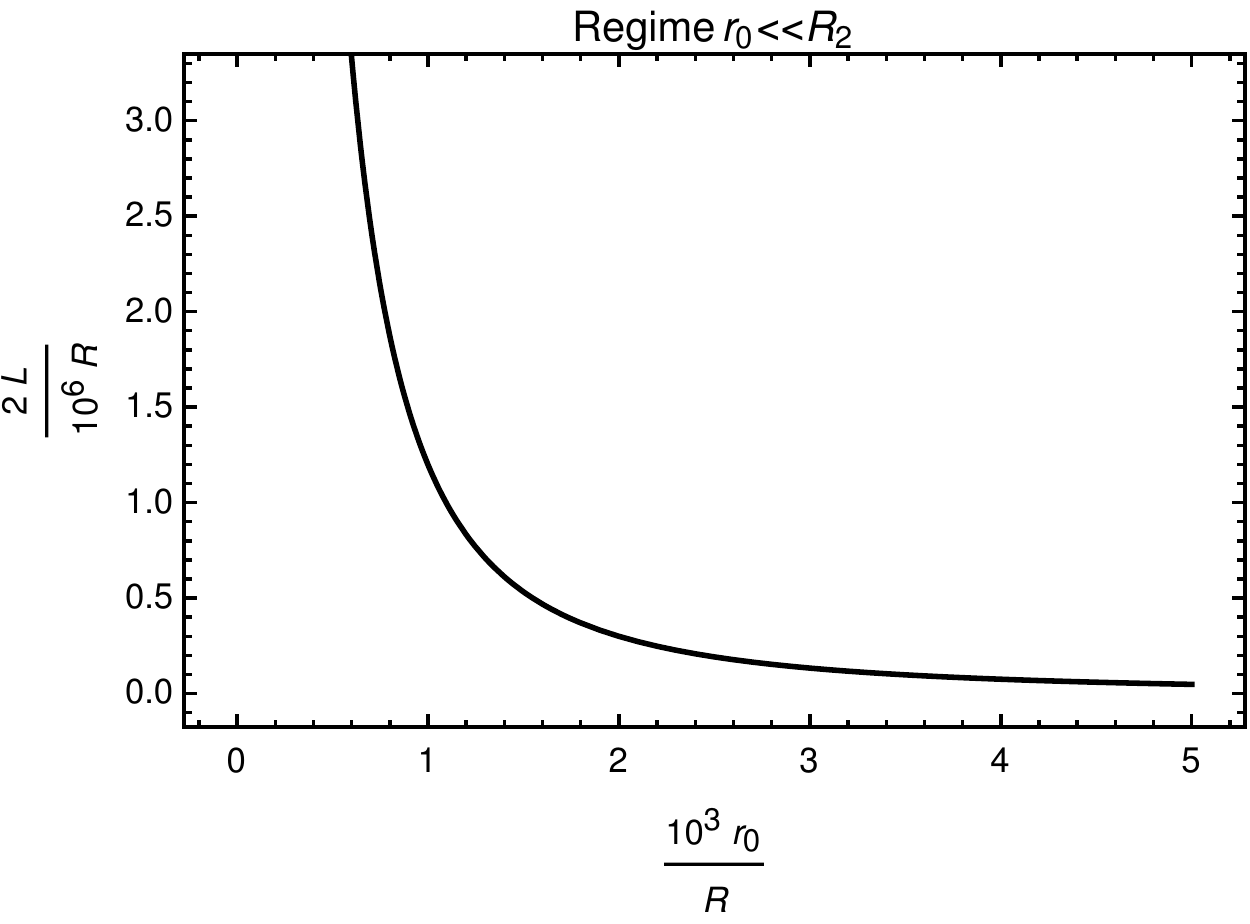}
 \caption{Distance $L/R_2$ between quarks in a M2-brane space vs. $r_0/R $. Here $r_1/R_2=10^4$, $r_0/R_2<5\times 10^{-3}$.}
 \label{fig:10}
\end{figure}

In this section we work  in the  geometric regime $r_0<<r_1$ which corresponds to very massive quarks and with $r_0<<R_2$ which means that we are in the near horizon geometry which is approximately AdS$_4$. 
First we plot in figure \ref{fig:10} the distance $L/R_2$, eq. (\ref{3.1}), against  $r_0/R_2$. We can notice from this plot 
that $L$ has a monotonic decreasing behaviour as $r_0$ is increased.

\begin{figure}[h]
 \centering
 \includegraphics[width=8cm,height=6cm]{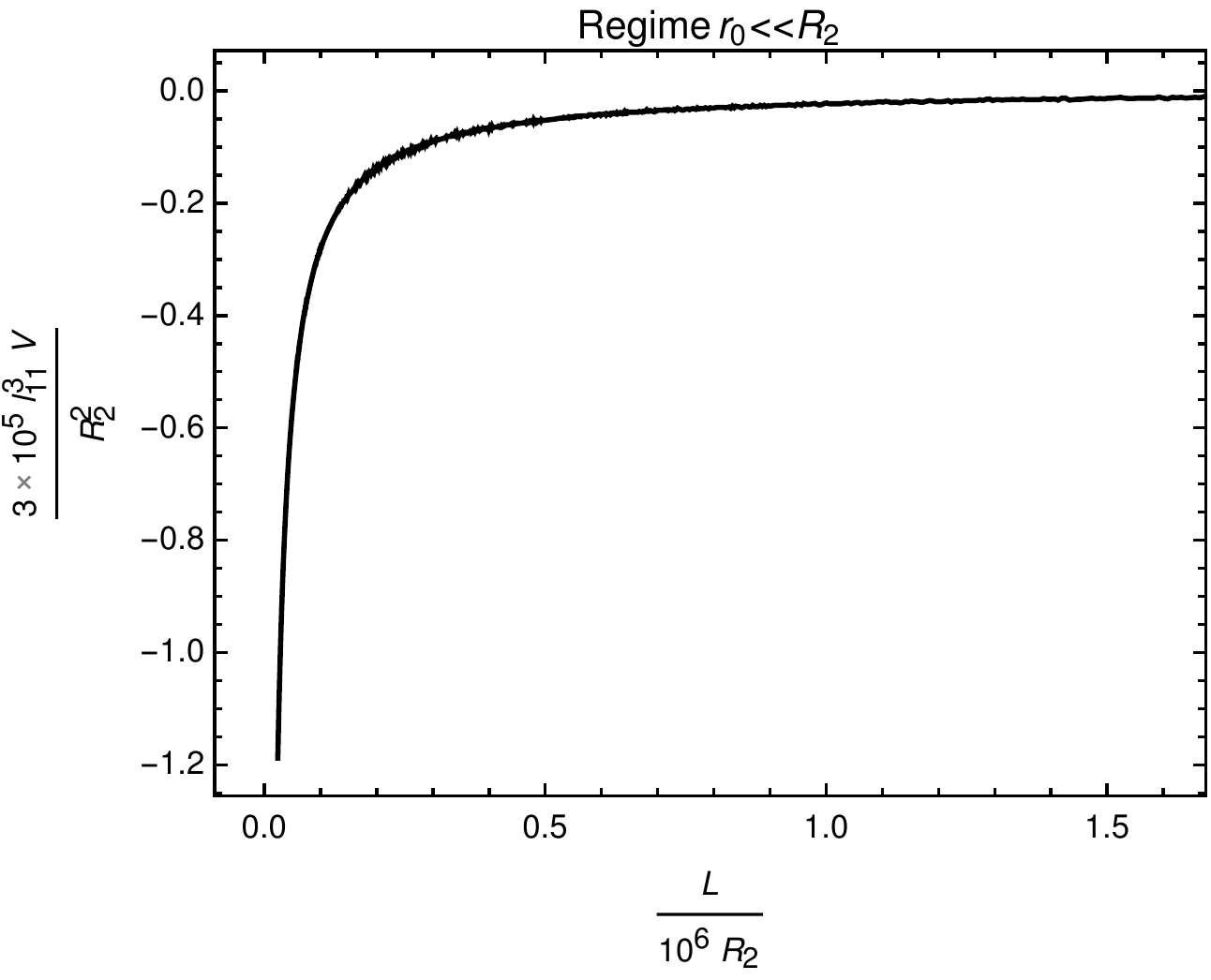}
 \caption{Potential $l_{11}^3V/R_2^2$ between quarks in a M2-brane space vs. $L/R_2$. Here $r_1/R_2=10^4$, $r_0/R_2<5\times 10^{-3}$.}
 \label{fig:11}
\end{figure}

Next, we plot in figure \ref{fig:11} the potential interaction $l_{11}^3V/R_2^2$, eq.(\ref{3.2}), against the distance of the  quark-antiquark pair 
$L/R_2$, eq.(\ref{3.1}). As we can note from this plot, the potential interaction in this case turns out to have a
 Coulomb-like non-confining behaviour. This is in agreement with the result obtained in this same regime in ref. \cite{Quijada:2015tma} and with the fact that the dual field theory is conformal, since we are in the near horizon geometry which is approximately AdS$_4$. 

\subsection{Confining behaviour}

\begin{figure}[h]
 \centering
 \includegraphics[width=7cm,height=6cm]{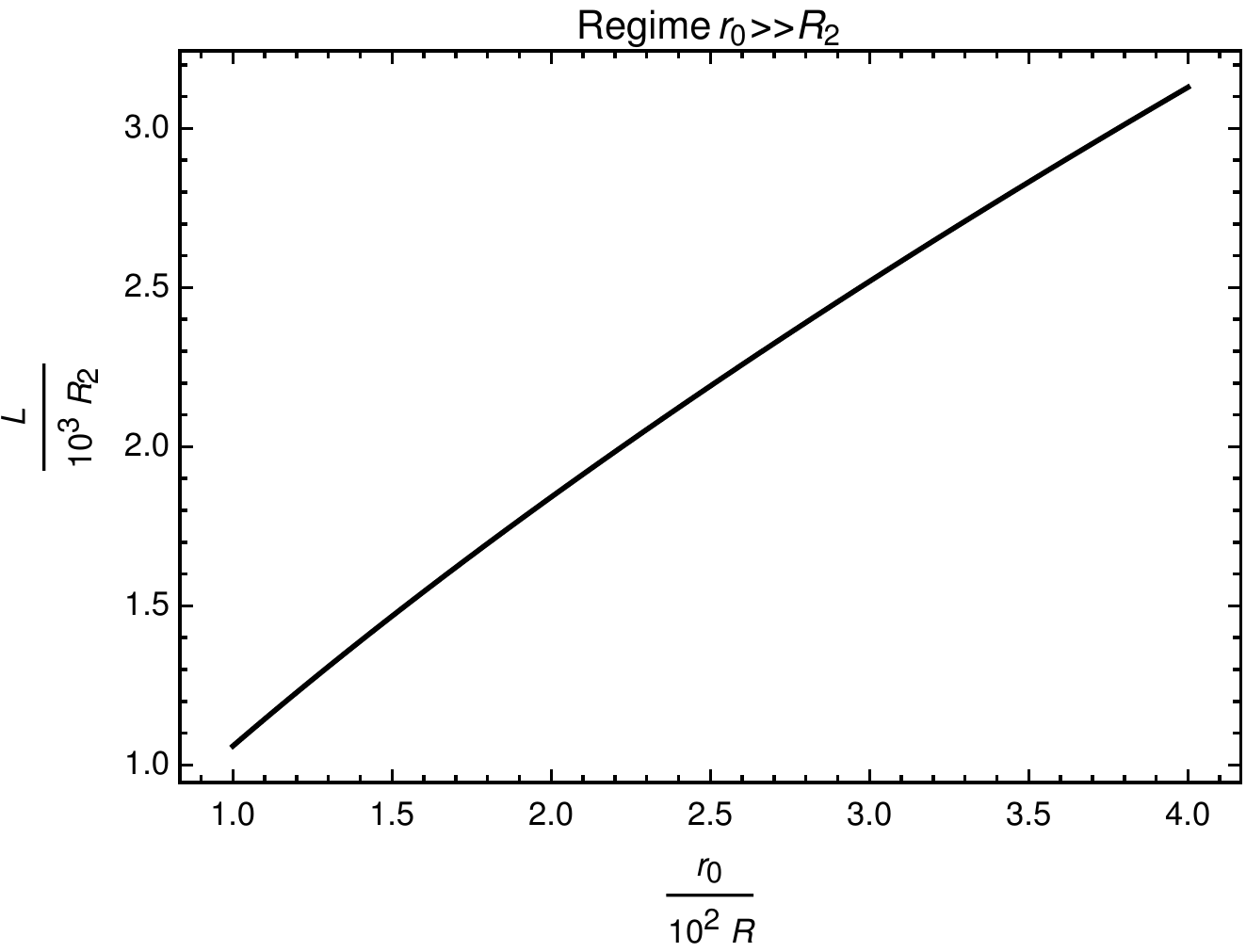}
 \caption{Distance $L/R_2$ between quarks in a M2-brane space vs. $r_0/R_2$. Here $r_1/R=10^4$ and $r_0/R_2>10^2$.}
 \label{fig:12}
\end{figure}

Here we still work in the regime  $r_1>>r_0$, but with $r_0>>R_2$, which corresponds to the region far from the horizon which is approximately a flat space. Now we plot in figure \ref{fig:12} the rationalised distance 
$L/R_2$, eq.(\ref{3.1}), against $r_0/R_2$. From this plot we can see that the distance $L$ has an increasing behaviour as $r_0$ is increased. Notice that this behaviour is almost linear.

\begin{figure}[h]
 \centering
 \includegraphics[width=7cm,height=6cm]{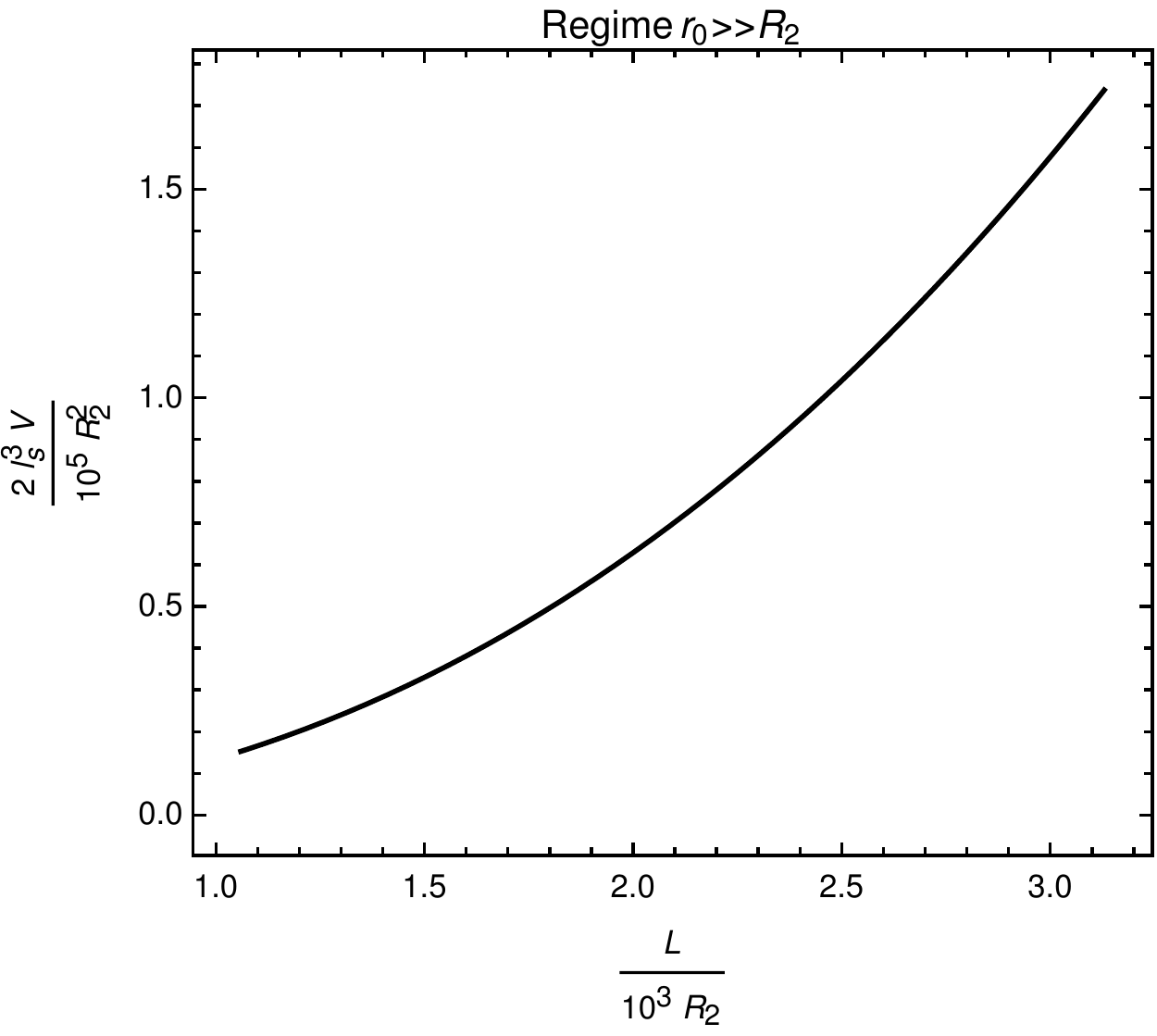}
 \caption{Potential $l_{11}^3V/R^2$ between quarks in a M2-brane space vs. $L/R_2$  for $r_0>>R_2$. Here $r_1/R=10^4$ and $r_0/R_2>10^2$.}
 \label{fig:13}
\end{figure}

Next, continuing in the same geometric regime, we plot  in figure \ref{fig:13} the potential $l_{11}^3V/R_2^2$, eq.(\ref{3.2}), against $L/R_2$, eq.(\ref{3.1}). We can
see from this plot that  potential interaction $V$ has  positive derivative, which means a confining behaviour. This behaviour is expected since we are working in the region far from the horizon of the M2-brane geometry where the dual field theory is non-conformal. 

\subsection{Deconfinement/Confinement transition}

In the last subsections we had a non-confining behaviour at the regime  $r_0<<R_2$ and a confining one at $r_0>>R_2$. So 
in this section we look for a transition behaviour at  a middle term regime $r_0\sim R_2$.

\begin{figure}[h]
 \centering
 \includegraphics[width=6cm,height=8cm]{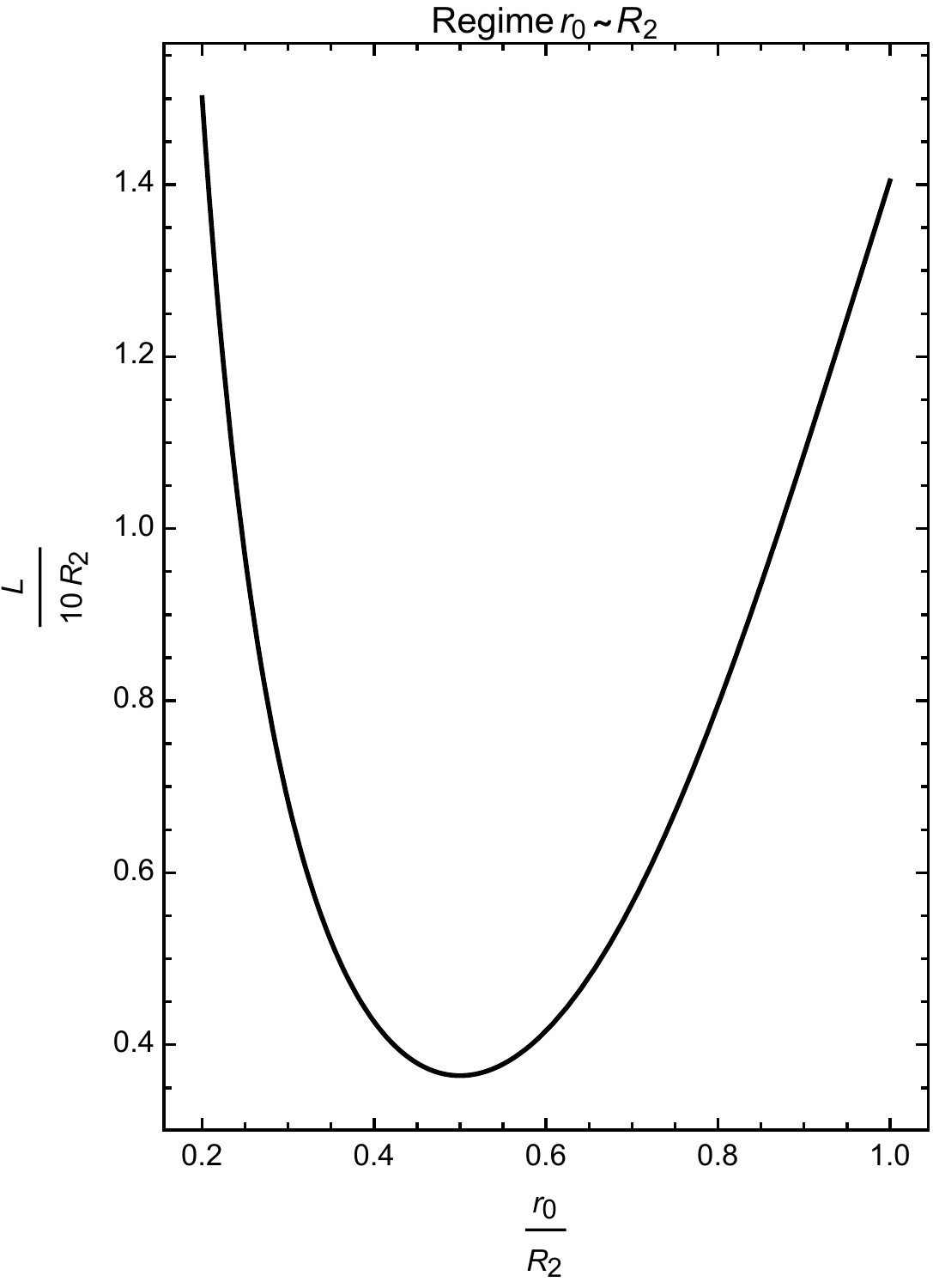}
 \caption{Distance $L/R_2$ between quarks in a M2-brane space vs. $r_0/R_2$. Here $r_1/R_2=10^4$, $r_0/R\leq1$ and $r*/R=0.500$.}
 \label{fig:14}
\end{figure}

First we plot the distance $L/R_2$ between quarks, eq.(\ref{3.1}),  against $r_0/R_2$, which is shown in figure \ref{fig:14}. From this plot we notice that there is a minimum at  $r_0=r*$. 

Also we can get this value $r*$ as a root of an equation that can be derived from (\ref{3.1}):
\begin{equation}\label{r*M2}
 \frac{d}{dr_0}L(r_0/R_2)=0\,. 
\end{equation}
Some solutions of this equation  are presented
on table \ref{table2} for different values of  $r_1/R_2$.
\begin{table}[h]
\centering
\caption{Some values of $r*/R_2$ in eq. \eqref{r*M2} for different values of $r_1/R_2$.}
\label{table2}
\begin{tabular}{|l|l|ll}
\cline{1-2}
$r*/R_2$ & $r_1/R_2$ &   &  \\ \cline{1-2}
0.500       & $1\times 10^4$              &  &   \\ \cline{1-2}
0.503      &  $8\times 10^3$        &  &  \\ \cline{1-2}
0.510       &    $4\times 10^3$        &  &  \\ \cline{1-2}
0.524      & $1\times 10^3$          &  &  \\ \cline{1-2}
\end{tabular}
\end{table}

\begin{figure}[h]
 \centering
 \includegraphics[width=7cm,height=8cm]{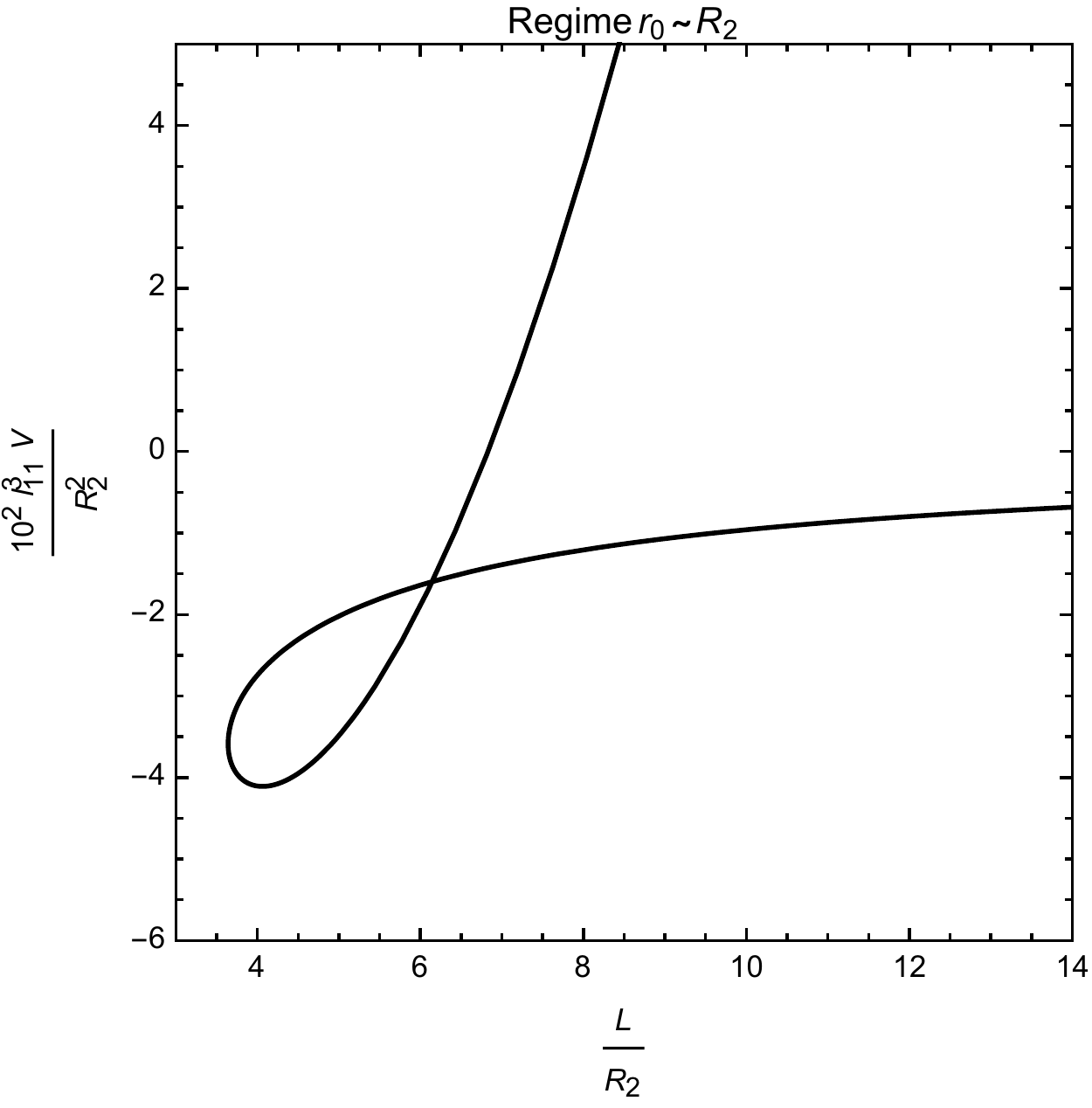}
 \caption{Potential   $l_{11}^3 V/R^2_2$ between quarks in a M2-brane space vs  $L/R_2$. }
 \label{fig:15}
\end{figure}

Next we plot in figure \ref{fig:15} the potential $l_{11}^3 V/R^2_2$, eq.(\ref{3.2}), against the distance  $L/R_2$, eq.(\ref{3.1}).
We notice from this plot that there are two branches: the inferior one corresponding to a non-confining Coulomb-like potential and 
the superior one corresponding to a confining potential. Also, from our analysis of the last plots, we can  conclude that  for  $r_0<r*$  the potential is a non-confining one while for $r_0>r*$  the  potential  is a confining one.

\section{M5-brane}

Now we analyse the confinement behaviour of a quark-antiquark pair using the MRY method in the 11-dimensional SUGRA background space generated by 
$N$ coincident M5-branes. The metric solution is \cite{Review2,Review1}:
\begin{equation}
 ds^2_{\text{M5}}=\left(1+\frac{R_5^3}{r^3}\right)^{-1/3}dx_6^2+\left(1+\frac{R_5^3}{r^3}\right)^{2/3}(dr^2+r^2d\Omega^2_4),
\end{equation}
where $R_5$ is a constant given by $R_5=(\pi N l_{11}^3)^{1/3}$.

Acoording to \cite{Quijada:2015tma}, the distance separation and the potential interaction of a pair of quarks are given by:
\begin{equation}\label{L5}
  L=2r_0\int_{1}^{r_1/r_0}dy\frac{(1+\epsilon/y^3)^{1/2}}{(y^2-1)^{1/2}}
\end{equation}
\begin{equation}\label{V5}
 V=\frac{r_0^2}{\pi l_{11}^3}\int_1^{r_1/r_0}dy\frac{y^2(1+\epsilon/y^3)^{1/2}}{(y^2-1)^{1/2}}-2m_q
\end{equation}
where  $r_0$ (- $r_1$) is the minimum(-maximum) value of coordinate $r$  of the string-like object obtained from dimensional reduction and  $\epsilon=R_5^3/r_0^3$.
Following ref. \cite{kinar} we can compute the quark mass:
\begin{equation}
2m_q= \frac{1}{\pi l_{11}^3}\int_0^{r_1}dr\, r\, \sqrt{1+\left(\frac{R_5}{r}\right)^3}\,,
\end{equation}
which is divergent in the limit $r_1\to\infty$.

\subsection{Non-confining behaviour}

\begin{figure}[h]
 \centering
 \includegraphics[width=8cm,height=7cm]{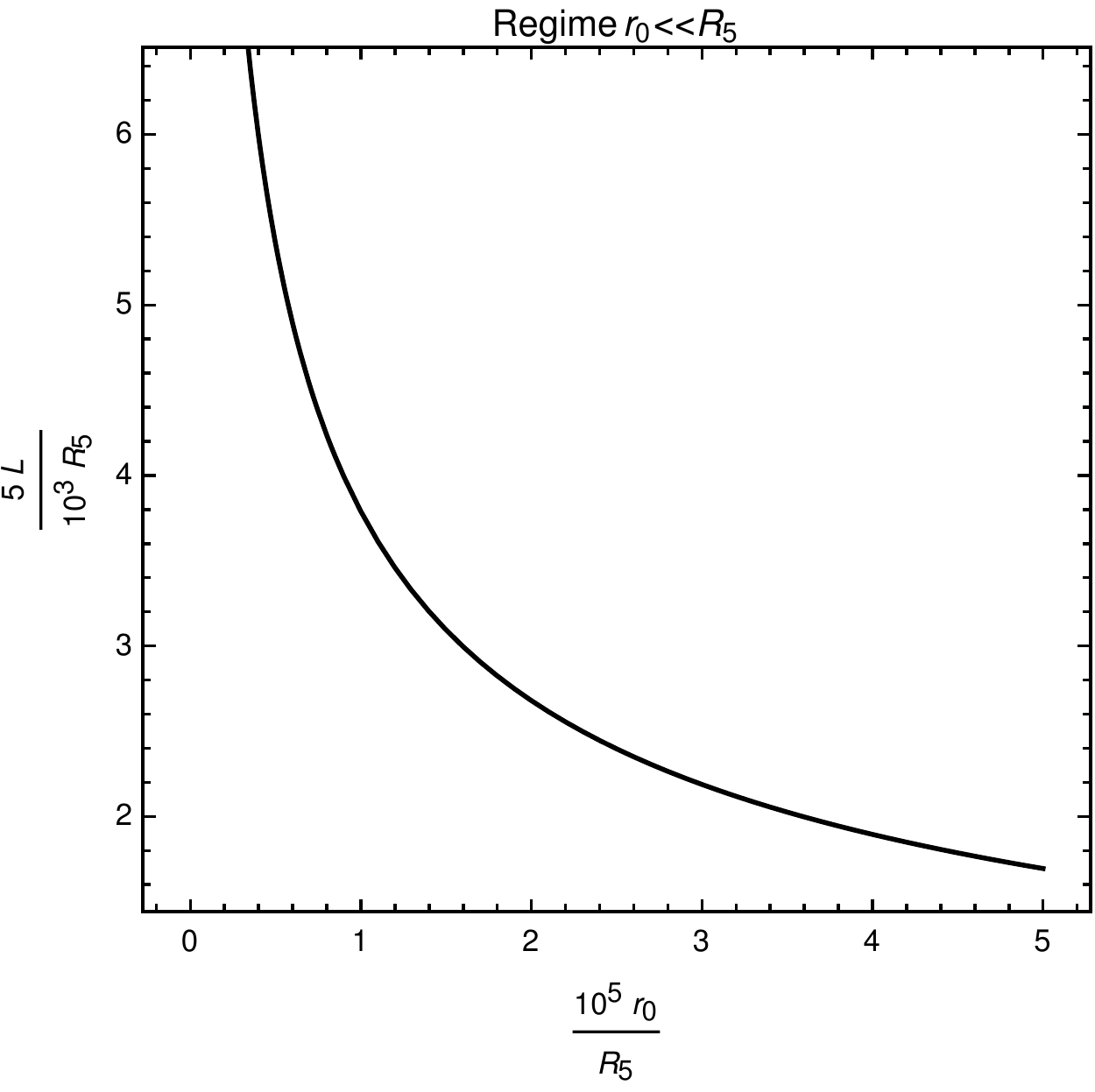}
 \caption{Distance  $L/R_5$ between quarks  vs. $r_0/R_5$. Here $r_1/R_5=10^2$, $r_0/R_5<5\times 10^{-5}$.}
 \label{fig:17}
\end{figure}

We work here  in the regime $r_1>>r_0$ which means that the quarks are very massive and with $r_0<<R_5$ corresponding the region near horizon which in this case is approximately an AdS$_7$ geometry. 
For this regime the distance $L/R_5$ between the pair of quarks, eq.(\ref{L5}), against
$r_0/R_5$ is plotted in figure \ref{fig:17}. This plot shows  that the distance $L$ has a 
monotonic decreasing behaviour  as $r_0$ is increased.

\begin{figure}[h]
 \centering
 \includegraphics[width=8cm,height=7cm]{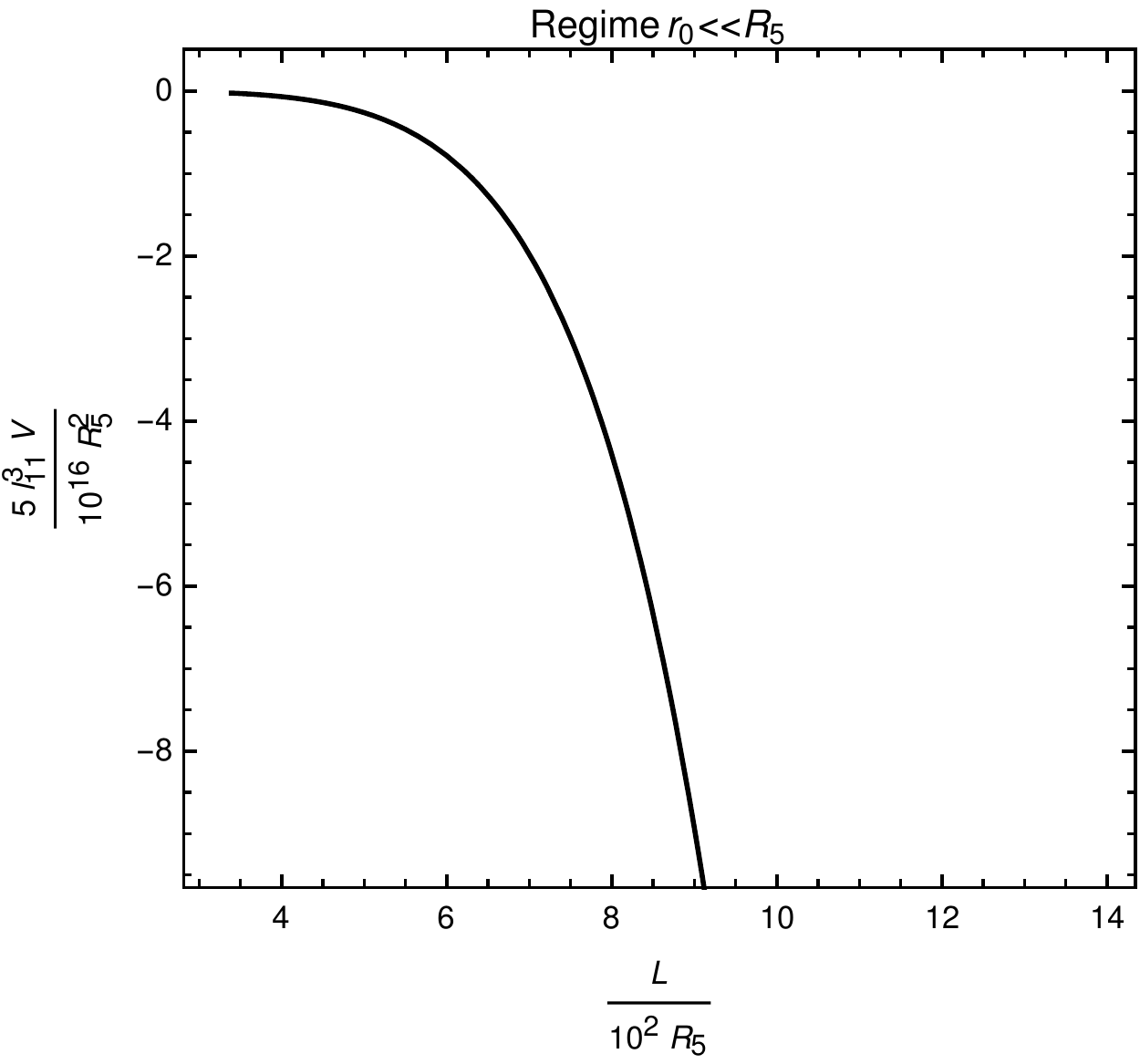}
 \caption{Potential $l_{11}^3V/R_5^2$  between quarks in a M5-brane space vs. the distance $L/R_5$. Here $r_1/R_5=10^2$ and $r_0/R_5<5\times 10^{-5}$.}
 \label{fig:18}
\end{figure}

Next we plot in figure \ref{fig:18}  the potential interaction $l_{11}^3V/R_5^2$, eq.(\ref{V5}), against the distance
separation between quarks $L/R_5$, eq.(\ref{L5}). We can see from this plot that the potential $V$ shows a non-confining behaviour: it has a negative slope and it goes to minus infinite as $L$  increases. This is the expected behaviour since we are in the near horizon region where the metric is aproximately an AdS$_7$ compatible with a conformal field theory. 

\subsection{Confining behaviour}

\begin{figure}[h]
 \centering
 \includegraphics[width=7cm,height=8cm]{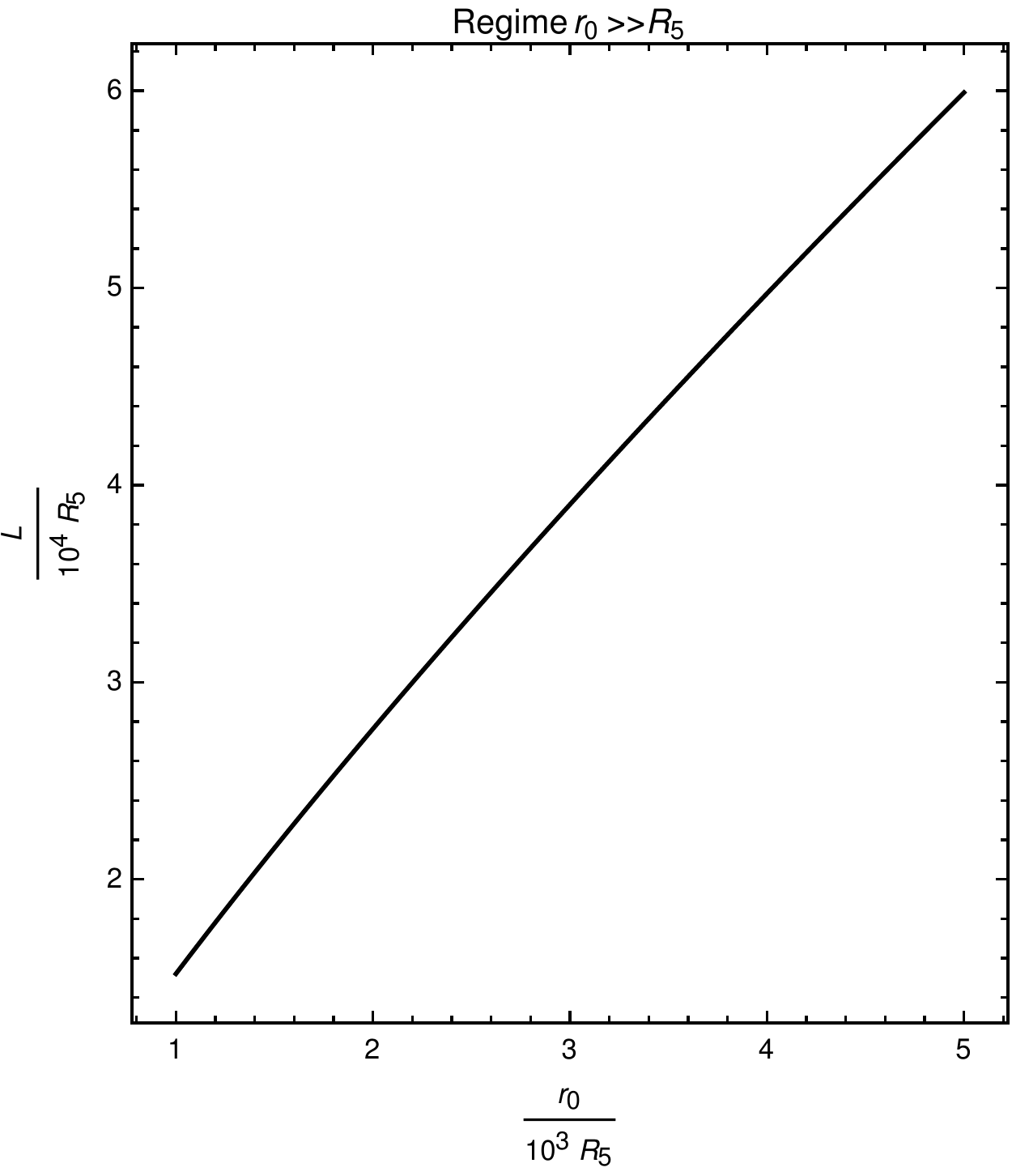}
 \caption{Distance  $L/R_5$  between quarks in a M5-brane space vs.  $r_0/R_5$. Here $r_1/R_5=10^6$ and $r_0/R_5>10^3$.}
 \label{fig:18a}
\end{figure}

In this subsection we still work  in the regime $r_1>>r_0$ but with $r_0>>R_5$ (far from the horizon) which corresponds to an approximately flat space geometry. In  
this regime we plot in figure \ref{fig:18a} the distance separation $L/R_5$, eq.(\ref{L5}), against $r_0/R_5$.
We can see from this plot that $L$ shows an almost linear behaviour as  $r_0$ is increasing.

\begin{figure}[h]
 \centering
 \includegraphics[width=8cm,height=8cm]{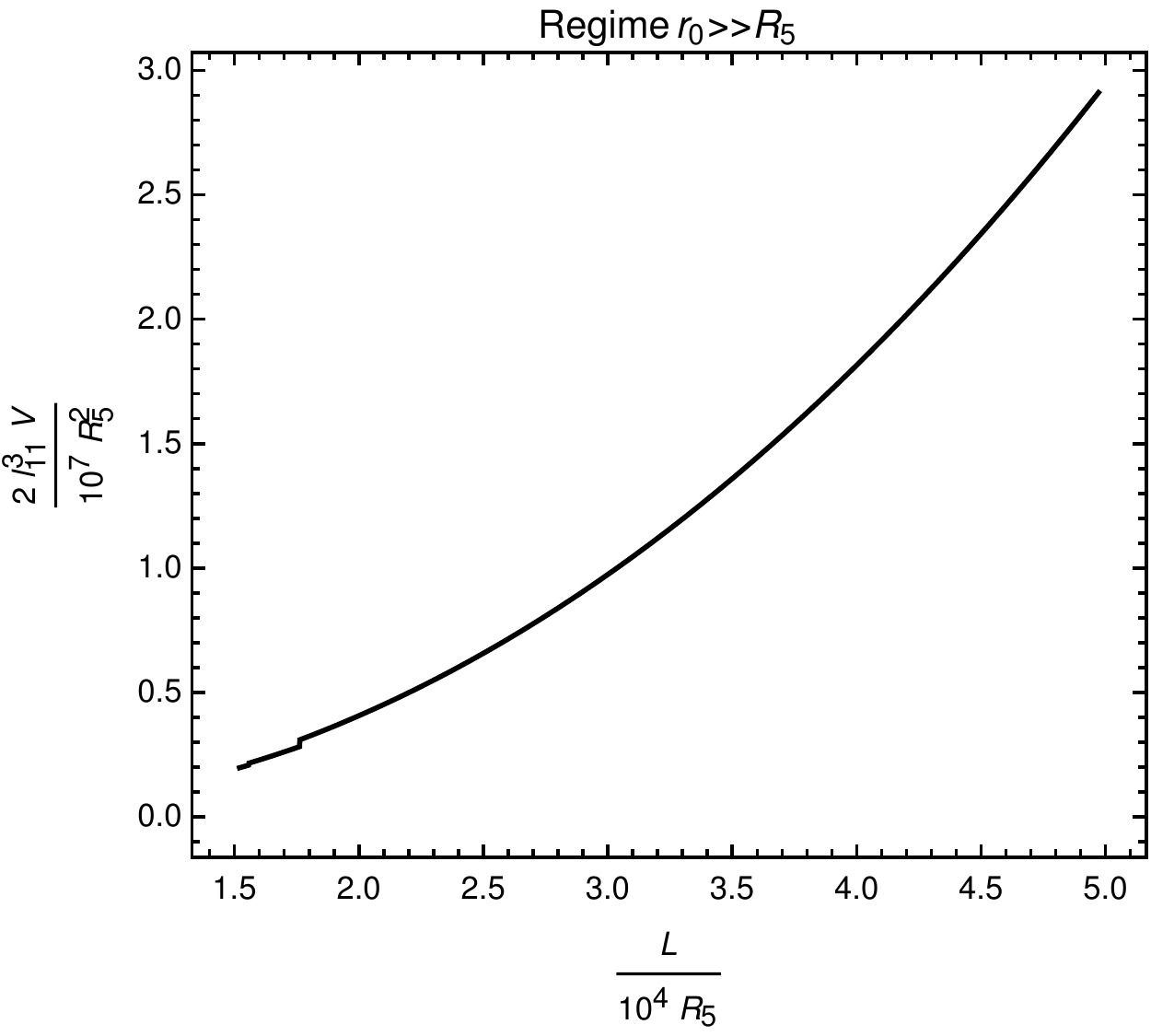}
 \caption{Potential interaction $l_{11}^3V/R_5^2$ between quarks in a M5-brane space vs. the distance $L/R_5$. Here $r_1/R_5=10^6$ and $r_0/R_5>10^3$.}
 \label{fig:18b}
\end{figure}

Next we plot in figure \ref{fig:18b} the
 potential   interaction $l_{11}^3V/R_5^2$ between the pair of quarks, eq.(\ref{V5}), against the distance between quarks $L/R_5$, eq.(\ref{L5}). From this plot we can notice that the potential $V$ shows a confining behaviour: it has a positive slope  as $L$ is increased. This behaviour is expected since we are working in the region far from the brane which approaches asymptotically a flat space so that the dual field theory is no longer conformal. 
 
\subsection{Deconfinement/Confinement transition}

\begin{figure}[h]
 \centering
 \includegraphics[width=7cm,height=6cm]{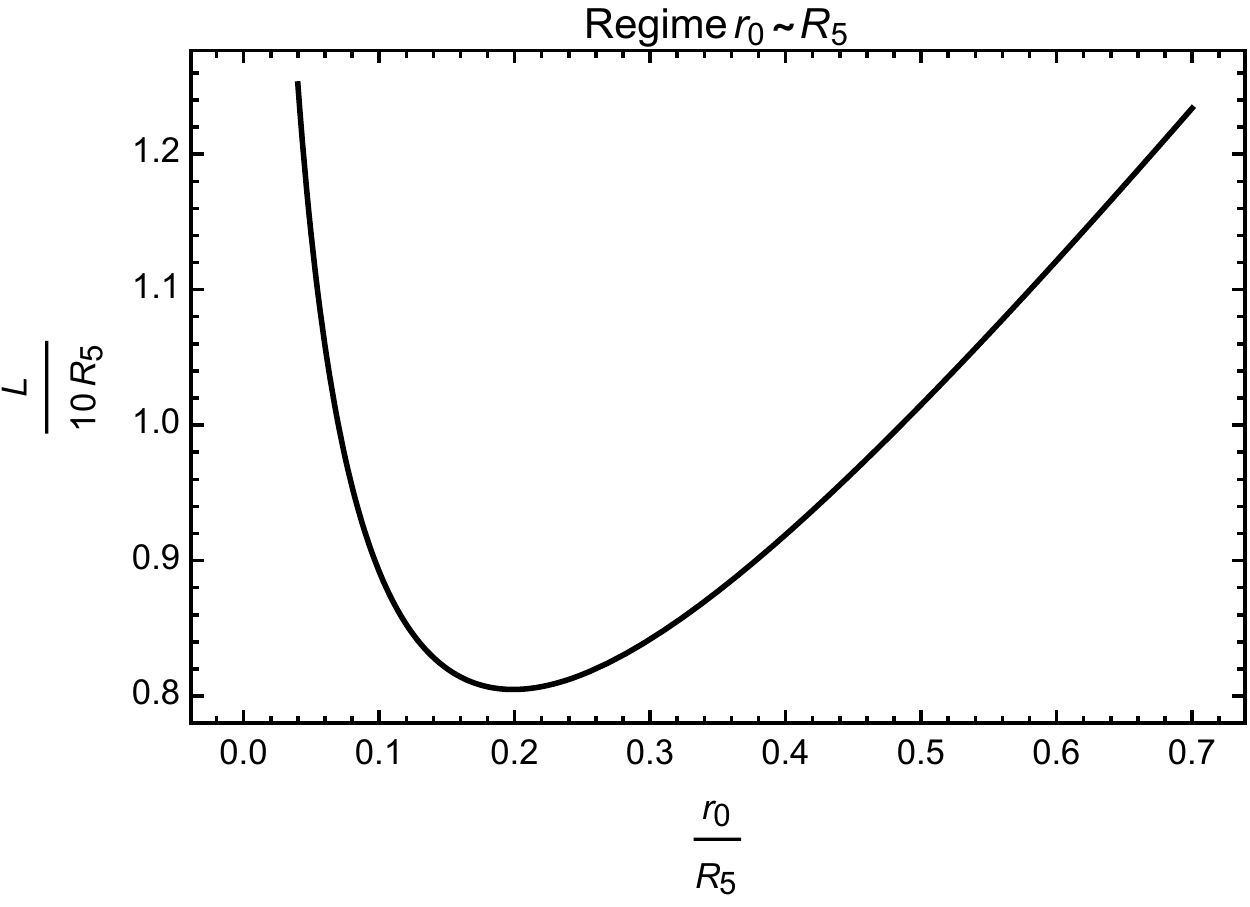}
 \caption{Distance $L/R_5$ between quarks in a M5-brane space vs.  $r_0/R_5$. Here $r_1/R_5=10^3$, $r_0/R_5<0.7$ and 
 $r*/R_5=0.19$.}
 \label{fig:19}
\end{figure}

In the last subsections we found non-confining potential behaviour at $r_0<<R_5$ and confining potential behaviour at $r_0>>R_5$. 
In this section we work in the regime $r_0\sim R_5$  and look for a confinement/deconfinement transition. 
First we plot in \ref{fig:19} the distance separation between quarks $L/R_5$, eq.(\ref{L5}), against $r_0/R_5$. From this plot we can notice
that there is a minimum at the position $r_0=r*$.

We can also get $r*$ as a root of a equation that is obtained  deriving  equation (\ref{L5}):
\begin{equation}\label{r*M5}
 \frac{d }{dr_0}L(r_0/R_5)=0\,.
\end{equation}
Some solutions of  this equation are shown in table \ref{table3} for some values of $r_1/R_5$.
\begin{table}[h]
\centering
\caption{Some values of $r*/R_5$ in eq. \eqref{r*M5} for different values of $r_1/R_5$.}
\label{table3}
\begin{tabular}{|l|l|ll}
\cline{1-2}
$r*/R_5$ & $r_1/R_5$ &   &  \\ \cline{1-2}
0.19     & $1\times 10^3$             &  &   \\ \cline{1-2}
0.21     & $5\times 10^2$    &  &  \\ \cline{1-2}
0.25     & $1 \times 10^2$     &  &  \\ \cline{1-2}
0.42     & $1\times 10^1$          &  &  \\ \cline{1-2}
\end{tabular}
\end{table}

\begin{figure}[h]
 \centering
 \includegraphics[width=7cm,height=6cm]{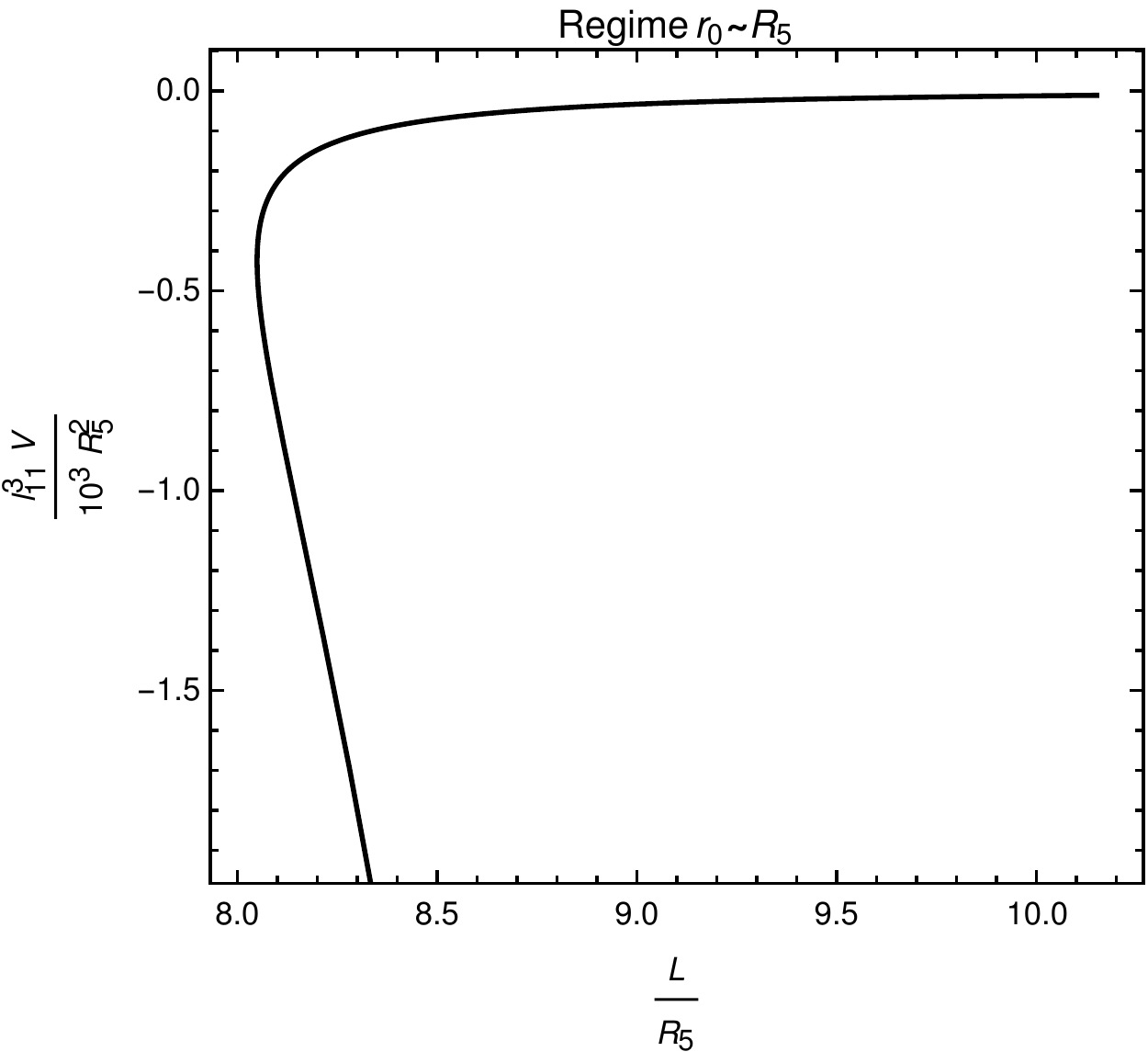}
 \caption{Potential interaction $l_{11}^3V/R_5^2$ between quarks in a M5-brane space vs.  $L/R_5$. Here $r_1/R_5=10^3$, $r_0/R_5<0.5 $ and 
 $r*/R_5=0.19$. }
 \label{fig:21}
\end{figure} 

\begin{figure}[h]
 \centering
 \includegraphics[width=7cm,height=6cm]{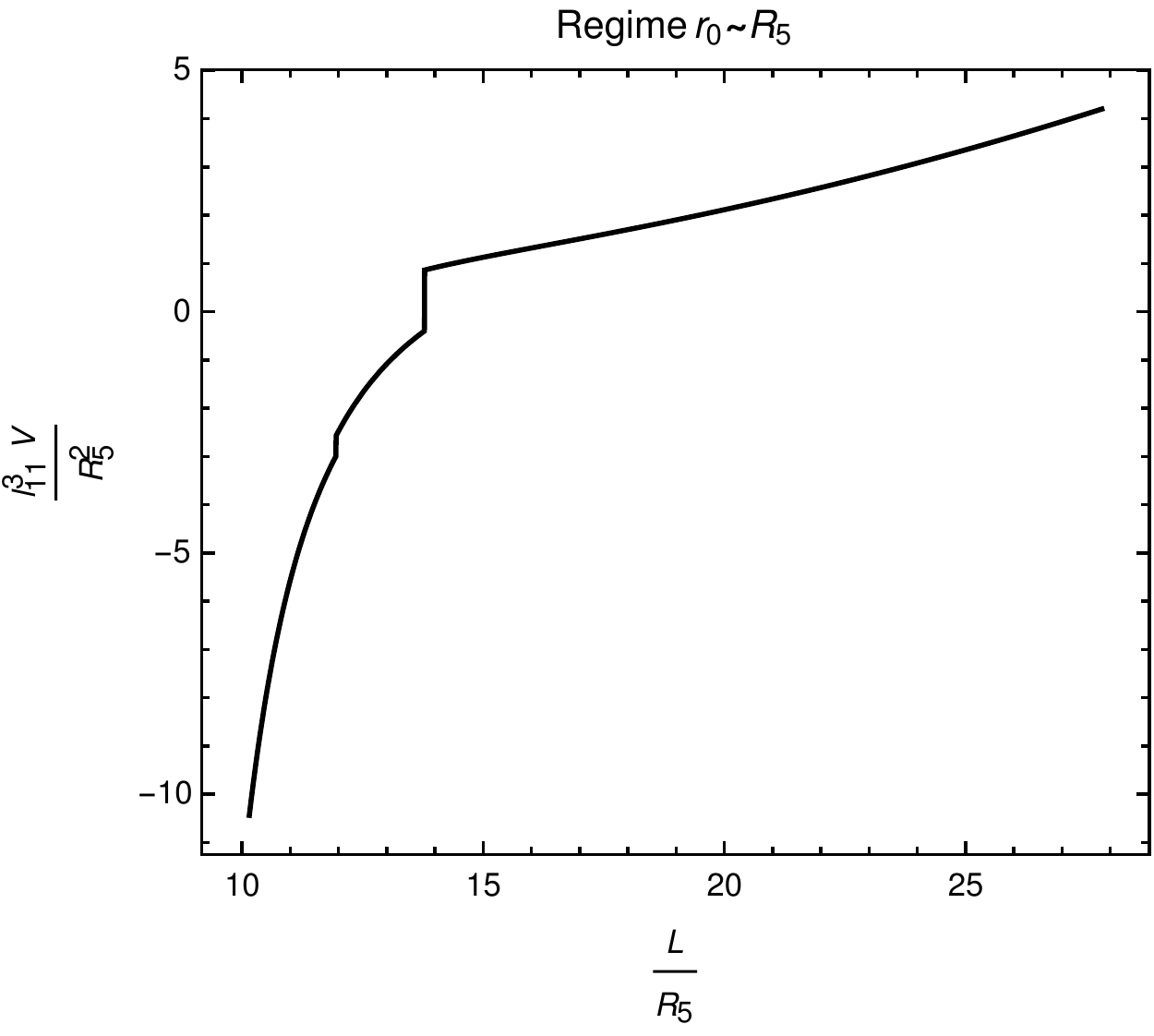}
 \caption{Potential interaction $l_{11}^3V/R_5^2$ between quarks in a M5-brane space vs.  $L/R_5$. Here $r_1/R_5=10^2$, $0.5<r_0/R_5$ and 
 $r*/R_5=0.19$. }
 \label{fig:22}
\end{figure}

Finally we plot in figures \ref{fig:21} and \ref{fig:22} the potential interaction $l_{11}^3V/R_5^2$, eq.(\ref{V5}), against the distance
separation between quarks $L/R_5$, 
eq.(\ref{L5}).
From these plots we can see  that we have two branches: 
The superior one in figure \ref{fig:21}, which continuation for larger values of $L/R_5$ is amplified in figure \ref{fig:22}, corresponds to a 
confining potential interaction, since we can observe that it has a positive derivative as $L$ increases. 
On the other side, the inferior one, 
that is just presented in figure \ref{fig:21}, corresponds to a non-confining potential.

Also we can conclude from these plots that  values with  $r_0<r*$  correspond to  a non-confining behaviour, and values with $r_0>r*$ correspond to a  confining behaviour.

\section{Conclusions}

We have analysed the Wilson loops for D3-, M2-, and M5-brane backgrounds using the MRY approach. 
As was discussed previously in refs. \cite{henrique,Quijada:2015tma} these backgrounds imply confining and non-confining quark-antiquark potentials depending on the geometric regime considered. We investigated here these situations further and mainly the transition between these two confinement behaviours. 

In general for the three geometries that we have studied, we notice that as the distance separation $L/R_i$ has a monotonic decreasing behaviour with  $r_0$, one finds a non-confining potential interaction. This situation occurs at the regime $r_1>>r_0$ and $r_0<<R_i$, which corresponds to heavy quark masses in AdS geometries ($R_i$ assumes the values $R$, $R_2$, and $R_5$ for the geometries D3-, M2-, and M5-branes, respectively). 

On the other hand, when the distance separation $L/R_i$ is a monotonic increasing function of $r_0$, one finds a   confining potential interaction. This situation occurs at the regime $r_1>>r_0$ and $r_0>>R_i$, which corresponds to heavy quark masses in flat space geometries. This confining behaviour can be understood looking at the metric in the region far from the brane. In this case the metric approaches a flat spacetime so that the dual field theory is no longer conformal. This situation is analogous to a string in flat space which mimics the flux tube model of QCD showing confinement. 

We found out that the confinement/deconfinement transition occurs at a point $r*$ in the regime $r_0\sim R_i$ for the D3, M2, and M5-brane backgrounds. The point $r*$ is where the non-monotonic  $L$ distance function of $r_0$ is a minimum. 
The value of $r*$ depends on $r_1$ and $R_i$ and we have tabulated possible values in tables \ref{my-label}, \ref{table2} and \ref{table3}, for each geometry. This situation occurs at the regime $r_1>>r_0$ (heavy quark) and corresponds to a transition between the AdS and flat space geometries. All these situations were analysed at zero temperature, so that the nature of the transitions are purely geometrical and not thermodynamical. 

It would be interesting to analyse if this discussion can be extended to other Wilson loop configurations where $1/N$ corrections are present \cite{Farquet:2014bda,Gomis:2006sb, Gomis:2006im, Buchbinder:2014nia}.

\acknowledgments We would like to thank C. Hoyos for interesting discussions at the {\it Strings at Dunes} conference in Natal, Brazil, 2016,  where a previous version of this work was presented. We would like also to thank CNPq - Brazilian agency - for financial support.  


\end{document}